\documentclass{article}
\usepackage{braket, amsmath, amssymb, graphicx, tabu, tikz, subcaption}
\usepackage[colorlinks = true]{hyperref}

\DeclareMathOperator{\diag}{diag}

\newtheorem{theorem}{Theorem}
\newtheorem{example}{Example}
\newtheorem{lemma}{Lemma}

\title{Phase Squeezing of Quantum Hypergraph States}
\author{Ramita Sarkar$^1$, Supriyo Dutta$^2$\thanks{Corresponding author. Email: \texttt{dosupriyo@gmail.com}}, Subhashish Banerjee$^3$, Prasanta K. Panigrahi$^1$ \vspace{.25cm}\\
$^1$ Department of Physical Sciences \\ Indian Institute of Science Education and Research Kolkata \\ Mohanpur, Nadia, West Bengal, India - 741246. \vspace{.25cm}\\
$^2$ Centre for Theoretical Studies \\ Indian Institute of Technology Kharagpur \\ Kharagpur, West Bengal, India - 721302. \vspace{.25cm}\\
$^3$ Department of Physics and IDRP-QIC\\ Indian Institute of Technology Jodhpur \\ NH 62 Nagaur Road, Karwar, Jodhpur Rajasthan, India - 342037.} 
\date{}

\begin{document}
	
	\maketitle
	
	\begin{abstract}
		Corresponding to a hypergraph $G$ with $d$ vertices, a quantum hypergraph state is defined by $\ket{G} = \frac{1}{\sqrt{2^d}}\sum_{n = 0}^{2^d - 1} (-1)^{f(n)} \ket{n}$, where $f$ is a $d$-variable Boolean function depending on the hypergraph $G$, and $\ket{n}$ denotes a binary vector of length $2^d$ with $1$ at $n$-th position for $n = 0, 1, \dots (2^d - 1)$. The non-classical properties of these states are studied. We consider annihilation and creation operator on the Hilbert space of dimension $2^d$ acting on the number states $\{\ket{n}: n = 0, 1, \dots (2^d - 1)\}$. The Hermitian number and phase operators, in finite dimensions, are constructed. The number-phase uncertainty for these states leads to the idea of phase squeezing. We establish that these states are squeezed in the phase quadrature only and satisfy the Agarwal-Tara criterion for non-classicality, which only depends on the number of vertices of the hypergraphs. We also point out that coherence is observed in the phase quadrature.
	\end{abstract}

	\section{Introduction}
		 
		The quantum graph states, also called the cluster states \cite{raussendorf2003measurement}, are well-studied quantum states which are used in different quantum information theoretic tasks \cite{nielsen2006cluster}. The quantum hypergraph states \cite{rossi2013quantum, qu2013encoding} are a generalization of these states. There is a one-to-one correspondence between the set of $n$-qubit hypergraph states and the set of $n$-variable Boolean functions \cite{dutta2018boolean}. In recent years they have been utilized in quantum error correction \cite{balakuntala2017quantum, wagner2018analysis} and quantum blockchain \cite{banerjee2020quantum}. Quantum optics provides a prominent platform for the physical implementation of quantum information theoretic tasks \cite{knill2001scheme,nielsen2004optical,himadri,sandhya}. For instance, optical squeezing is applied for carrying out the algorithms in quantum cryptography \cite{gehring2015implementation}. Hence, investigating the non-classical properties of quantum hypergraph states from the perspective of quantum optics would be pertinent.

		The quantum hypergraph state is a family of finite dimensional quantum states. The recent developments in quantum state engineering, computing and communication stimulate the production and manipulation of finite dimensional quantum states. Non-classicality of these quantum states is an important facet of investigations in quantum optics \cite{samya,priya1,priya2}. Different finite dimensional quantum states are considered in this context, for instance, the binomial states \cite{fu1996generalized, mandal2019generalized}, negative binomial states \cite{agarwal1992negative}, hypergeometric states \cite{fu1997hypergeometric}. Most of these analysis are focused on their constructions as well as the possible occurrence of various nonclassical effects exhibited by them. Introducing a graph-theoretic or a combinatorial framework in the investigation sheds further light in this direction \cite{bibhas}. Entanglement of quantum states is a non-classical property. Graph theory has been emploed to the problem of detecting entanglement \cite{dutta2016bipartite, haddadi2019efficient, akhound2019analyzing}. Entanglement in quantum hypergraph states are well-studied in literature, see, for instance, \cite{dutta2019permutation} and references therein.
		
		Here, we study the nonclassical behavior of quantum hypergraph states. An analytical study of non-classicality for finite dimensional states turns out to be difficult \cite{buvzek1992coherent, glauber1963quantum, miranowicz1994coherent}. Given any hypergraph with $d$ vertics, the general analytical form of these states is spanned by $2^d$ number states $\ket{0}, \ket{1}, \ket{2}, \dots \ket{2^d - 1}$ in a $2^d$ dimensional Hilbert space $\mathcal{H}^{2^d}$. The physical variables, which are the expectation values of Hermitian operators are evaluated in this basis. Therefore, these mean values depend parametrically on the number of vertices $d$. The number and phase operators are seen to be non-commutative for these states. Although, the quantum hypergraph states allow non-zero number-phase uncertainty we detect squeezing, a prominent aspect of non-classicality, in phase quadrature only. Recall that, the existence of a Hermitian phase operator of the harmonic oscillator is a long-standing open problem in quantum mechanics \cite{dirac1925fundamental, dirac1927quantum, scully1991quantum}. We follow the Pegg-Barnett formalism in the construction of the phase operator in finite dimensional space \cite{pegg1988unitary, pegg1989phase, buvzek1992coherent}. Interestingly, the Agarwal-Tara criterion for non-classicality \cite{agarwal1992nonclassical}, which is associated to the higher order moments of number operator, holds for these states. Coherence is another important facet of quantumness \cite{baumgratz2014quantifying}. Evolution of coherence has been studied in the context of open quantum systems \cite{samya}, as well as in sub-atomic systems \cite{dixit2019study}. Coherence is studied, here, for various classes of hypergraphs in both number and phase basis.

		The description of quantum hypergraph states and their relevance to quantum information theory is discussed in section \ref{motivation}. The preliminary concepts of annihilation and creation operators acting on finite dimensional Hilbert spaces are also presented. In section \ref{Squeezing in number and phase quadrature}, the study of squeezing in quantum hygregraph states is made. The construction of Hermitian phase operator for the finite dimensional Hilbert space as well as the number-phase uncertainty relation is also developed. We establish that these states are squeezed in phase quadrature only. The degree of squeezing is calculated for different types of hypergraphs. Section \ref{discussions_on_Agarwal_Tara_criterion} discusses the Agarwal-Tara criterion for non-classicality, which depends on the number of vertices of the hypergraphs. Coherence for various classes of hypergraphs in both number and phase basis is discussed in section \ref{coherence}. We then make our Conclusions. In appendix \ref{appendix_A} we present a few essential tools of linear algebra as well as properties of Toeplitz and circulant matrices, of relevance to our work. The appendix \ref{appendix_B} contains the expressions of higher order moments of the number operator, essential for the detailed calculation of the Agarwal-Tara criterion.	\\

	\section{The hypergraph states}\label{motivation}
		
		In combinatorics, a graph $G = (V(G), E(G))$ is a combination of a set of vertices $V(G)$ and a set of edges $E(G)$ \cite{west1996introduction}. Throughout this article, $d$ denotes the number of vertices in a graph or hypergraph with the vertex set $V(G) = \{1, 2, \dots d\}$. A simple edge is a set of two vertices $e = \{u, v\}$. A hypergraph $G = (V(G), E(G))$ is a generalization of graphs, such that, $E(G)$ contains at least one hyperedge $e$, that is a set of more than two vertices \cite{bretto2013hypergraph}.
		
		To define a graph state or a hypergraph state we assign a $\ket{+} = \frac{1}{\sqrt{2}} \begin{bmatrix} 1 \\ 1 \end{bmatrix}$ state corresponding to every vertex of the graph or the hypergraph $G$. Now, for edge $\{u, v\}$ we apply a $2$-qubit controlled-NOT gate on the states corresponding to $u$ and $v$. Similarly, for a hyperedge containing $r$ vertices we apply an $r$-qubit controlled-NOT gate on the states corresponding to the vertices in the hyperedge. It generates the following quantum state, known as the quantum hypergraph state:
		\begin{equation}\label{optical_hypergraph_state}
			\ket{G} = \frac{1}{\sqrt{2^d}}\sum_{n = 0}^{2^d - 1} (-1)^{f(n)} \ket{n},
		\end{equation}
		where $\ket{n} = (0, 0, \dots 0, 1 ((n + 1)\text{-th position}), 0, \dots 0)^t$ is a basis vector in $\mathcal{H}^{2^d}$, and $f : \{0,1\}^{d} \rightarrow \{0, 1\}$ is a Boolean function of $d$ variables depending on $G$. The explicit relation between $G$ and $f$ is discussed in \cite{dutta2018boolean}. In quantum information parlance, $\ket{n}$ is expressed as a $d$-qubit state which is a basis of $\mathcal{H}_2^{\otimes d}$, where $\mathcal{H}_2$ is the space generated by the basis vectors $\begin{bmatrix} 1 \\ 0 \end{bmatrix}$ and $\begin{bmatrix} 0 \\ 1 \end{bmatrix}$. Here, we neglect the multi-qubit structure. This enables the application of the annihilation and creation operators to a broad range of number states $\ket{n}$. As $\ket{G}$ is described by the state vector of multi-qubit hypergraph states, the present study could be expected  to have practical implications.
		
		\begin{example}
			A hypergraph $G = (V(G), E(G))$ with $V(G) = \{0, 1, 2, 3\}$ and $E(G) = \{(0, 1), (0, 2, 3), (1, 2, 3)\}$ is depicted in the figure \ref{example_hypergraph}. For generating its corresponding hypergraph state, we apply the multi-qubit CNOT gates on $\ket{+}^{\otimes 4}$ which are drawn in figure \ref{example_circuit}. The resultant quantum state is given by
			\begin{equation}
				\begin{split}
					& \ket{G} = \frac{1}{4}[1,  1,  1,  1,  1,  1,  1, -1,  1, -1,  1,  1,  1, -1,  1, -1]^t \\ 
					= & \frac{1}{4}\left[ \ket{0} + \ket{1} + \ket{2} + \ket{3} + \ket{4} + \ket{5} + \ket{6} - \ket{7} + \ket{8} - \ket{9} + \ket{10} + \ket{11} + \ket{12} - \ket{13} + \ket{14} - \ket{15} \right].
				\end{split}
			\end{equation}
			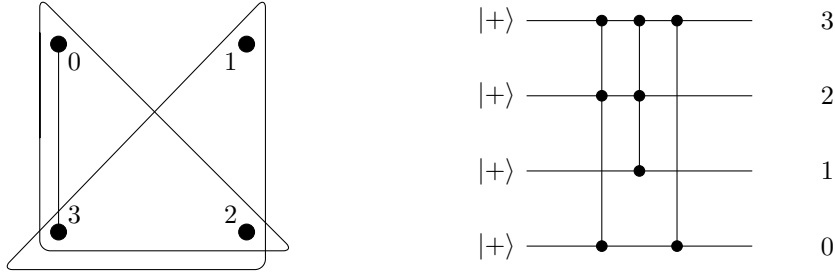
\begin{figure}
				\centering
				\begin{subfigure}{0.4\textwidth}
					\centering 
					\begin{tikzpicture}[scale = 2.5]
						\draw[fill] (0, 1) circle [radius = 1.2pt];
						\node[below right] at (0, 1) {$0$};
						\draw[fill] (1, 1) circle [radius = 1.2pt];
						\node[below left] at (1, 1) {$1$};
						\draw[fill] (1, 0) circle [radius = 1.2pt];
						\node[above left] at (1, 0) {$2$};
						\draw[fill] (0, 0) circle [radius = 1.2pt];
						\node[above right] at (0, 0) {$3$};
						\draw (0, 1)--(0,0);
						\draw [rounded corners] (-0.1, .5) -- (-0.1, 1.1) -- (-.1, -.1) -- (1.25, -.1) -- (-0.1, 1.25) -- (-0.1, .5);
						\draw [rounded corners] (1.1, .5) -- (1.1, -.2) -- (-.3, -.2) -- (1.1, 1.25) -- (1.1, .5);
					\end{tikzpicture}
					\caption{A hypergraph with four vertices $0, 1, 2$ and $3$ as well as an edge $(0, 3)$ and two hyperedges $(0, 2, 3)$ and $(1, 2, 3)$.}
					\label{example_hypergraph}
				\end{subfigure} 
				~
				\begin{subfigure}{0.4\textwidth}
					\centering 
					\begin{tikzpicture}[scale = 1]
						\draw (0, 0) -- (3, 0);
						\draw (0, 1) -- (3, 1);
						\draw (0, 2) -- (3, 2);
						\draw (0, 3) -- (3, 3);
						\node at (-.4, 0) {$\ket{+}$};
						\node at (-.4, 1) {$\ket{+}$};
						\node at (-.4, 2) {$\ket{+}$};
						\node at (-.4, 3) {$\ket{+}$};
						\node at (4, 0) {$0$};
						\node at (4, 1) {$1$};
						\node at (4, 2) {$2$};
						\node at (4, 3) {$3$};
						\draw[fill] (1, 3) circle [radius = 2pt];
						\draw[fill] (1, 0) circle [radius = 2pt];
						\draw[fill] (1, 2) circle [radius = 2pt];
						\draw (1, 0) -- (1, 3);
						\draw[fill] (1.5, 2) circle [radius = 2pt];
						\draw[fill] (1.5, 1) circle [radius = 2pt];
						\draw[fill] (1.5, 3) circle [radius = 2pt];
						\draw (1.5, 1) -- (1.5, 3);
						\draw[fill] (2, 0) circle [radius = 2pt];
						\draw[fill] (2, 3) circle [radius = 2pt];
						\draw (2, 0) -- (2,3);
					\end{tikzpicture}
					\caption{Quantum circuit for generating the hypergraph state corresponding to the hypergraph depicted in figure \ref{example_hypergraph}.}
					\label{example_circuit}
				\end{subfigure}
				\caption{A hypergraph and its corresponding quantum circuit}
			\end{figure}
		\end{example}
		
		In the Hilbert space $\mathcal{H}^{2^d}$ the creation and annihilation operators are represented by the following matrices 
		\begin{equation}
		a^\dagger = \begin{bmatrix}
		0 & 0 & 0 & \dots & 0 & 0 \\ \sqrt{1} & 0 & 0 & \dots & 0 & 0 \\ 0 & \sqrt{2} & 0 & \dots & 0 & 0 \\ 0 & 0 & \sqrt{3} & \dots & 0 & 0 \\ \vdots & \vdots & \vdots & \ddots & \vdots & \vdots \\ 0 & 0 & 0 & \hdots & \sqrt{2^d - 1} & 0
		\end{bmatrix} 
		~\text{and}~ 
		a = \begin{bmatrix}
		0 & \sqrt{1} & 0 & 0 & \dots & 0 \\ 0 & 0 & \sqrt{2} & 0 & \dots & 0 \\ 0 & 0 & 0 & \sqrt{3} & \dots & 0 \\ \vdots & \vdots & \vdots & \vdots & \ddots & \vdots \\ 0 & 0 & 0 & 0 & \dots & \sqrt{2^d - 1} \\ 0 & 0 & 0 & 0 & \dots & 0
		\end{bmatrix}.
		\end{equation}
		Matrix multiplication brings out the commutation relation between the annihilation and creation operator as \cite{buvzek1992coherent, miranowicz1994coherent}
		\begin{equation}\label{a_a_dagger_commutation}
			[a, a^\dagger] = aa^\dagger - a^\dagger a = I - 2^d\ket{2^d-1} \bra{2^d-1},
		\end{equation}
		which is different from the commutation relation $aa^\dagger - a^\dagger a = I$ in infinite dimensional Hilbert space \cite{popov1991photon, pegg1988unitary, pegg1989phase}. The annihilation operator $a$ acts on a number state $\ket{i}$ as $a\ket{i} = \sqrt{i}\ket{i - 1}$ for $i = 1, 2, \dots (2^d - 1)$, and $a\ket{0} = 0$, which is the all zero vector. Similarly, for the creation operator we have $a^\dagger\ket{i} = \sqrt{i+1}\ket{i + 1}$ for $i = 0, 1, \dots (2^d - 2)$ and $a^\dagger \ket{2^d - 1} = 0$. Note that, this assumption is different from the action of annihilation and creation operators on qubit states.

	\section{Squeezing in number and phase quadrature}\label{Squeezing in number and phase quadrature}
	
		We denote and define the average and the variance of an operator $\hat{A}$ with respect to the state $\ket{G}$ by $\langle \hat{A} \rangle = \braket{G | \hat{A} | G}$ and $\langle (\Delta \hat{A})^2 \rangle = \langle \hat{A}^2 \rangle - \langle \hat{A} \rangle^2$, respectively. Two operators $\hat{A}$ and $\hat{B}$ commute with respect to $\ket{G}$ if $\langle [\hat{A}, \hat{B}] \rangle = \braket{G | \hat{A} \hat{B} - \hat{B}\hat{A} | G} = 0$. If $\hat{A}$ and $\hat{B}$ do not commute we have the uncertainty relation \cite{hall2013quantum}
		\begin{equation}
			\langle (\Delta \hat{A})^2 \rangle \langle (\Delta \hat{B})^2 \rangle \geq \frac{1}{4} |\langle [\hat{A}, \hat{B}] \rangle|.
		\end{equation}
		We say that the variances of the operators $\hat{A}$ and $\hat{B}$ are squeezed if $\langle (\Delta \hat{A})^2 \rangle < \frac{1}{2} |\langle [\hat{A}, \hat{B}] \rangle|$ or $\langle (\Delta \hat{B})^2 \rangle < \frac{1}{2} |\langle [\hat{A}, \hat{B}] \rangle|$ \cite{wodkiewicz1985coherent}. The degree of squeezing in the quadrature of $\hat{A}$ and $\hat{B}$ are given by
		\begin{equation}
			\begin{split}
				S_{\hat{A}} = \frac{\langle (\Delta \hat{A})^2 \rangle - \frac{1}{2} |\langle [\hat{A}, \hat{B}] \rangle|}{\frac{1}{2} |\langle [\hat{A}, \hat{B}] \rangle|} ~\text{and}~ S_{\hat{B}} = \frac{\langle (\Delta \hat{B})^2 \rangle - \frac{1}{2} |\langle [\hat{A}, \hat{B}] \rangle|}{\frac{1}{2} |\langle [\hat{A}, \hat{B}] \rangle|},
			\end{split}
		\end{equation}
		respectively. It is observed that if $S_{\hat{A}} < 0$ then $S_{\hat{B}} > 0$ and  if $S_{\hat{B}} < 0$ then $S_{\hat{A}} > 0$.
		
		A number of squeezed states have been studied in the literature \cite{dodonov2002nonclassical}. Among them are states squeezed in position and momentum quadrature. The position and momentum operators are defined by $\hat{X} = \frac{1}{\sqrt{2}}(a+a^\dagger)$, and $\hat{P} = \frac{1}{i\sqrt{2}}(a-a^\dagger)$, respectively. Applying equation (\ref{a_a_dagger_commutation}) we can show that
		\begin{equation}
			\begin{split}\label{xp_com}
				[\hat{X}, \hat{P}] = & iI - 2^d i\ket{2^n-1}\bra{2^n-1} \\
				\text{or}~ \braket{[\hat{X}, \hat{P}]} = & i \braket{G|G} - 2^d i \braket{G|2^n-1} \braket{2^n-1| G} = 0.
			\end{split}
		\end{equation}
		Further we can prove that $\braket{[\hat{X}^k, \hat{P}^k]} = 0$ for any positive integer $k$. These calculations leads us to the conclusion that there is no uncertainty in the quadrature $\hat{X}$ and $\hat{P}$ for the hypergraph states.
		
		Considering the number and phase operators instead of $\hat{X}$ and $\hat{P}$ we observe an uncertainty relation for the hypergraph states \cite{barnett2007quantum, buvzek1992coherent}. Recall that in equation (\ref{optical_hypergraph_state}) we assume the states $\{\ket{n} : n = 0, 1, \dots, (2^d - 1)\}$ as the number states. Clearly, $\braket{m | n} = \delta_{m, n}$ and $\sum_{n = 0}^{2^d - 1} \ket{n} \bra{n} = I_{2^d}$. Now, the number operator is defined by $\hat{N} = \sum_{n = 0}^{2^d - 1} n \ket{n} \bra{n}$. The average of $\hat{N}$ is given by
		\begin{equation}
			\braket{\hat{N}} = \braket{G | \hat{N} |G } = \frac{1}{2^d} \sum_{n = 0}^{2^d - 1} (-1)^{f(n) + f(n)} n = \frac{(2^d - 1)2^d}{2 \times 2^d} = \frac{2^d - 1}{2}.
		\end{equation}
		Also,
		\begin{equation}
		\braket{\hat{N}^2} = \braket{G | \hat{N}^2 |G } = \frac{1}{2^d} \sum_{n = 0}^{2^d - 1} (-1)^{f(n) + f(n)} n^2 = \frac{(2^d - 1)2^d(2^{d + 1} - 1)}{6 \times 2^d} = \frac{(2^d - 1)(2^{d + 1} - 1)}{6}.
		\end{equation}
		Therefore, 
		\begin{eqnarray}
			\braket{(\Delta \hat{N})^2} = \braket{\hat{N}^2} - \braket{\hat{N}}^2 = \frac{(2^d - 1)(2^{d + 1} - 1)}{6} - \left(\frac{2^d - 1}{2}\right)^2 = \frac{(2^d - 1)(2^d + 1)}{12}.
		\end{eqnarray}
		Note that, $\braket{\hat{N}}$ and $\braket{(\Delta \hat{N})^2}$ depend only on the number of vertices in the hypergraph $G$.  
		
		For $m = 0, 1, \dots (2^d - 1)$, the phase $\theta_m$ is defined by $\theta_m = \theta_0 + 2 \pi \frac{m}{2^d}$, where we consider $\theta_0 = 0$, for simplicity. The phase states \cite{priyaphase1,priyaphase2} are defined by
		\begin{equation}\label{phase_state}
			\ket{\theta_m} = \frac{1}{\sqrt{2^d}} \sum_{n = 0}^{2^d - 1} \exp \left(\iota \theta_m  n\right) \ket{n}.
		\end{equation}
		Note that, in the above expression $\exp \left(\iota \theta_m  n\right)$ are the $2^d$-th complex root of unity. It can be proved that $\braket{\theta_i| \theta_j} = \delta_{i, j}$ and $\sum_{i = 0}^{2^d - 1} \ket{\theta_i} \bra{\theta_i} = I$.
		
		The phase operator \cite{perinova,agarwalsrinivasan,sbphasediss,sbphaseqnd} is defined by $\hat{P} = \sum_{m = 0}^{2^d - 1} \theta_m \ket{\theta_m} \bra{\theta_m}$, which can be expanded in the number basis as: 
		\begin{equation}\label{phase_operator_in_number_state}
			\begin{split}
				\hat{P} & = \sum_{q = 0}^{2^d - 1}  \sum_{r = 0}^{2^d - 1} p_{qr} \ket{\theta_q} \bra{\theta_r} ~\text{where}~  p_{qr} = \begin{cases} \theta_r & ~\text{for}~ q = r \\ 0 & ~\text{otherwise} \end{cases} \\
				& = \sum_{q = 0}^{2^d - 1}  \sum_{r = 0}^{2^d - 1} p_{qr} \left[ \frac{1}{\sqrt{2^d}} \sum_{k = 0}^{2^d - 1} \exp \left(\iota \theta_q  k\right) \ket{k} \right] \left[ \frac{1}{\sqrt{2^d}} \sum_{l = 0}^{2^d - 1} \exp \left(-\iota \theta_r l\right) \bra{l} \right] \\
				& = \frac{1}{2^d} \sum_{q = 0}^{2^d - 1}  \sum_{r = 0}^{2^d - 1} p_{qr} \left[ \sum_{k = 0}^{2^d - 1} \sum_{l = 0}^{2^d - 1} \exp \left(\iota \theta_q  k -\iota \theta_r l \right) \ket{k}\bra{l} \right] \\
				& = \frac{1}{2^d} \sum_{r = 0}^{2^d - 1} \frac{2 \pi r}{2^d} \left[ \sum_{k = 0}^{2^d - 1} \sum_{l = 0}^{2^d - 1} \exp \left(\iota \theta_r (k - l) \right) \ket{k}\bra{l} \right] ~\text{[putting}~ q = r ~\text{and}~ \theta_r = \frac{2 \pi r}{2^d}] \\
				& = \frac{2 \pi}{4^d} \sum_{k = 0}^{2^d - 1} \sum_{l = 0}^{2^d - 1} \left[ \sum_{r = 0}^{2^d - 1} r \exp \left(\iota \theta_r (k - l) \right) \right] \ket{k}\bra{l}\\
			\end{split} 
		\end{equation}
		\begin{equation*} 
			\begin{split} 
				~\text{or}~ \hat{P} & = \frac{2 \pi}{4^d} \sum_{k = 0}^{2^d - 1} \sum_{l = 0}^{2^d - 1} P_{k, l} \ket{k}\bra{l}, ~\text{where} P_{k, l} = \sum_{r = 0}^{2^d - 1} r \exp \left(\iota \theta_r (k - l) \right).
			\end{split}
		\end{equation*}
		The above calculation suggests that $\hat{P}$ is a Hermitian circulant matrix which is proved in appendix \ref{appendix_A}.  The expectation value of $\hat{P}$ is
		\begin{equation}
			\braket{G | \hat{P} | G} = \sum_{m = 0}^{2^d - 1} \theta_m \braket{G | \theta_m} \braket{\theta_m | G} = \sum_{m = 0}^{2^d - 1} \theta_m \overline{ \braket{\theta_m | G}} \braket{\theta_m | G} = \sum_{m = 0}^{2^d - 1} \theta_m |\braket{\theta_m | G}|^2.
		\end{equation}
		Also,
		\begin{equation}
			\braket{G | \hat{P}^2 | G} = \sum_{m = 0}^{2^d - 1} \theta_m^2 \braket{G | \theta_m} \braket{\theta_m | G} = \sum_{m = 0}^{2^d - 1} \theta_m^2 \overline{ \braket{\theta_m | G}} \braket{\theta_m | G} = \sum_{m = 0}^{2^d - 1} \theta_m^2 |\braket{\theta_m | G}|^2.
		\end{equation}
		Therefore the variance of $\hat{P}$ is given by
		\begin{equation}\label{variance_of_phase_operator}
			\langle (\Delta \hat{P})^2 \rangle = \braket{G | \hat{P}^2 | G} - \braket{G | \hat{P} | G}^2 = \sum_{m = 0}^{2^d - 1} \theta_m^2 |\braket{\theta_m | G}|^2 - \left| \sum_{m = 0}^{2^d - 1} \theta_m |\braket{\theta_m | G}|^2 \right|^2.
		\end{equation}
		It can be observed that 
		\begin{equation}
			\begin{split} 
				\braket{\theta_m| G} = & \frac{1}{\sqrt{2^d}} \sum_{n = 0}^{2^d - 1} \exp \left(-\iota \theta_m  n\right) \braket{G | n} = \frac{1}{\sqrt{2^d}} \sum_{n = 0}^{2^d - 1} (-1)^{f(n)} \exp \left(-\iota \theta_m  n\right).
			\end{split} 
		\end{equation} 
		
		The number phase commutation operator $[\hat{N}, \hat{P}]$ is given by
		\begin{equation}\label{commutation_operator_in_number_basis}
			\begin{split}
				& [\hat{N}, \hat{P}] = \hat{N}\hat{P}  - \hat{P}\hat{N} \\
				& = \frac{2 \pi}{4^d} \left[ \left( \sum_{k = 0}^{2^d - 1} \sum_{l = 0}^{2^d - 1} n_{k, l} \ket{k}\bra{l} \right) \left( \sum_{k = 0}^{2^d - 1} \sum_{l = 0}^{2^d - 1} P_{k, l} \ket{k}\bra{l} \right) - \left( \sum_{k = 0}^{2^d - 1} \sum_{l = 0}^{2^d - 1} P_{k, l} \ket{k}\bra{l} \right) \left( \sum_{k = 0}^{2^d - 1} \sum_{l = 0}^{2^d - 1} n_{k, l} \ket{k}\bra{l} \right) \right] \\
				& ~\text{where}~ n_{k, l} = \begin{cases} l & ~\text{for}~ k = l \\ 0 & ~\text{otherwise} \end{cases} \\
				& = \frac{2 \pi}{4^d} \left[ \sum_{k = 0}^{2^d - 1} \sum_{l = 0}^{2^d - 1} k P_{k, l} \ket{k}\bra{l} - \sum_{k = 0}^{2^d - 1} \sum_{l = 0}^{2^d - 1} P_{k, l} l \ket{k}\bra{l} \right] = \frac{2 \pi}{4^d} \sum_{k = 0}^{2^d - 1} \sum_{l = 0}^{2^d - 1} \left( k - l \right) P_{k, l} \ket{k}\bra{l}.
			\end{split}
		\end{equation}
		The matrix $[\hat{N}, \hat{P}]$ is a skew-Hermitian Toeplitz matrix. The proof can be seen in appendix \ref{appendix_A}. As $[\hat{N}, \hat{P}]$ is a skew-Hermitian matrix, its diagonal entries are all zero. Also, by expanding $[\hat{N}, \hat{P}]$ we observe that $\sum_{l = 1}^{2^d - 1} |a_{0,l}| = \max\{\sum_{k \neq l} |a_{k, l}|: ~\text{for}~ k = 0, 1, \dots (2^d - 1)\}$, where $a_{k, l} = \frac{2 \pi}{4^d}( k - l) P_{k, l} $ where $k \neq l$ are the off diagonal entries of $[\hat{N}, \hat{P}]$ in a particular row. Therefore if $\lambda$ is an eigenvalue of $[\hat{N}, \hat{P}]$, the Gershgorin circle theorem suggests that $|\lambda| \leq \sum_{l = 1}^{2^d - 1} |a_{0,l}|$. This is shown in appendix \ref{appendix_A}. Further, as $[\hat{N}, \hat{P}]$ is a Toeplitz matrix, all the rows of $[\hat{N}, \hat{P}]$ are determined by the entries of its $k = 0$-th row, which is given by a row vector
		\begin{equation}
			\frac{2 \pi}{4^d} \sum_{l = 0}^{2^d - 1} \left(- l \right) P_{0, l} \bra{l} = - \frac{2 \pi}{4^d} \sum_{l = 0}^{2^d - 1} \sum_{r = 0}^{2^d - 1} lr \exp \left(-i \frac{2 \pi l r}{2^d}\right) \bra{l}.
		\end{equation}
		Recall that $\exp \left(-i \frac{2 \pi l r}{2^d}\right)$ is a $2^d$-th root of unity. Now, summing over the absolute values of the individual entries we find
		\begin{equation}
			\sum_{l = 1}^{2^d - 1} |a_{0, l}| = \frac{2 \pi}{4^d} \sum_{l = 0}^{2^d - 1} \sum_{r = 0}^{2^d - 1} \left| lr \exp \left(-i \frac{2 \pi l r}{2^d}\right) \right| = \frac{2 \pi}{4^d} \sum_{l = 0}^{2^d - 1} \sum_{r = 0}^{2^d - 1} lr = \frac{2 \pi (2^d - 1)^2}{4^{d + 1}}.
		\end{equation}
		Therefore, for any eigenvalue $\lambda$ of $[\hat{N}, \hat{P}]$ we find $|\lambda| \leq \frac{2 \pi (2^d - 1)^2}{4^{d + 1}}$. As $[\hat{N}, \hat{P}]$ is a skew-Hermitian matrix, $\iota [\hat{N}, \hat{P}]$ is Hermitian. Also $\braket{\iota [\hat{N}, \hat{P}]} = \iota \braket{[\hat{N}, \hat{P}]}$. Taking the absolute value we have $|\braket{[\hat{N}, \hat{P}]}| = |\iota \braket{[\hat{N}, \hat{P}]}|$. Using the idea of Rayleigh quotient, see appendix \ref{appendix_A}, we have
		\begin{equation}
			\lambda_{\min} \leq |\braket{[\hat{N}, \hat{P}]}| \leq \lambda_{\max},
		\end{equation}
		for any possible state. The degree of squeezing with respect to the number operator $\hat{N}$ is defined by
		\begin{equation}
			S_{\hat{N}} = \frac{ \braket{(\Delta \hat{N})^2}- \frac{1}{2}|\braket{[\hat{N}, \hat{P}]}| }{\frac{1}{2}|\braket{[\hat{N}, \hat{P}]}|}.
		\end{equation}
		Now, 
		\begin{equation}
			\braket{(\Delta \hat{N})^2} - \frac{1}{2}|\braket{[\hat{N}, \hat{P}]}| \geq \braket{(\Delta \hat{N})^2} - \frac{1}{2} |\lambda_{\max}| \geq \frac{(2^d - 1)(2^d + 1)}{12} - \frac{\pi (2^d - 1)^2}{4^{d + 1}} \geq 0,
		\end{equation}
		for any $d$. Therefore, the quantum hypergraph state $\ket{G}$ has no squeezing with respect to the number operator. This observation can be precisely written as follows:
		\begin{theorem}
			There are quantum hypergraph states which are squeezed in the phase quadrature only.
		\end{theorem} 
		But, not all hypergraph states are squeeed in phase quadrature. The hypergraph in figure \ref{example_hypergraph} is a negative example, for which $\braket{(\Delta \hat{P})^2} = 3.4312$, $\braket{(\Delta \hat{N})^2} = 21.25$ and $\frac{1}{2}|\braket{[\hat{N}, \hat{P}]}| = 1.8624$.
	
		Now, we shall discuss about the hypergraph states with phase squeezing. The degree of squeezing with respect to the phase operator $\hat{P}$ is defined by 
		\begin{equation}\label{squeezing_in_phase_quadrature}
			S_{\hat{P}} = \frac{ \braket{(\Delta \hat{P})^2}- \frac{1}{2}|\braket{[\hat{N}, \hat{P}]}| }{\frac{1}{2}|\braket{[\hat{N}, \hat{P}]}|}.
		\end{equation}
		Equation (\ref{variance_of_phase_operator}) suggests that $S_{\hat{P}}$ depends on the choice of the hypergraph states $\ket{G}$. However, we have not yet succeeded in obtaining an explicit expression of $S_{\hat{P}}$ depending on the hypergraph $G$. In addition, for any integer $d \geq 2$ there are $2^{2^d}$ quantum states of dimension $2^d$. Hence, numerical evaluation of $S_{\hat{P}}$ for all hypergraphs with $d$ vertices is a very tedious task. Therefore, we choose a very special class of hypergraphs for our investigations. We have the following numerical observations:
		\begin{enumerate}
			\item 
				Let the hypergraph $G$ have $d$ vertices and no hyperedge. Therefore the corrsponding hypergraph state is $\ket{G} = \frac{1}{\sqrt{2^d}}\sum_{n = 0}^{2^d - 1} \ket{n}$, which is a constant vector. From equation (\ref{phase_state}) it can be seen that
				\begin{equation}
					\braket{G | \theta_m} = \frac{1}{2^d} \sum_{n = 0}^{2^d - 1} \exp \left(\iota \theta_m n\right) = 0,
				\end{equation}
				as $\exp \left(\iota \theta_m n\right)$ are the complex $2^d$-th root of $1$. Putting this in equation (\ref{variance_of_phase_operator}) we find that the varience of $\hat{P}$ is zero. But $\braket{[\hat{N}, \hat{P}]}$ need not be zero. Hence, the degree of squeezing mentioned in equation (\ref{squeezing_in_phase_quadrature}) is $-1$, which is the minimum value of squeezing.
			\item 
				Let $G$ be a hypergraph with number of vertices $d > 3$ and exactly one hyperedge containing all the vertices. Then $\ket{G}$ is squeezed in the phase quadrature. Also, the values of $S_{\hat{P}}$ become asymptotically close to $-1$. The values are plotted in figure \ref{squeezing_of_connected_hypergraphs_with_single_hyperedge} and mentioned in table \ref{squeezing_of_connected_hypergraphs_with_single_hyperedge_table}. We depict some of these hypergraphs in figure \ref{connected_hypergraphs_with_single_edges}.
				\begin{figure} 
					\centering 
					\begin{subfigure}{.4 \textwidth}
						\includegraphics[scale = .45]{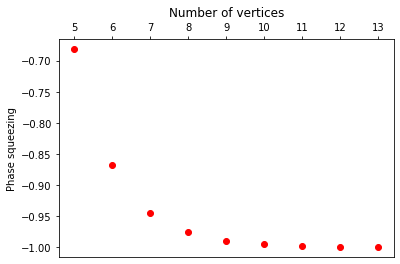}
						\caption{Squeezing of connected hypergraphs with single hyperedge} 
						\label{squeezing_of_connected_hypergraphs_with_single_hyperedge}
					\end{subfigure}
					\hspace{1cm}
					\begin{subfigure}{.4 \textwidth}
						\includegraphics[scale = .45]{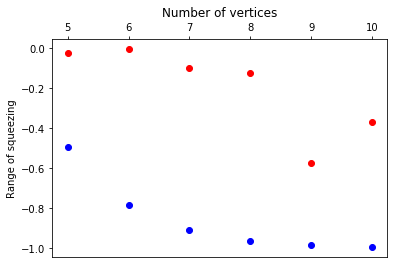}
						\caption{Maximum and minimum values of squeezing for $(d - 1)$-graphs. (Red and blue points represent the minimum and maximum)} 
						\label{maximum_and_minimum_squeezing_for_(d-1)_graphs_figure}
					\end{subfigure}
					\caption{Squeezing for different hypergraph states.}
				\end{figure} 
				\begin{table}
					\begin{tabular}{| l | l | l | l | l | l | l | l | l | l | l |}
						\hline
						$d$ & $4$ & $5$ & $6$ & $7$ &  $8$ & $9$ & $10$ & $11$ & $12$ & $13$ \\
						\hline
						$S_{\hat{P}}$ & $-0.2238$ & $-0.6817$ & $-0.8686$ & $-0.9449$ & $-0.9764$ & $-0.9898$ & $-0.9955$ & $-0.998$ & $-0.9991$ & $-0.9996$ \\
						\hline 
					\end{tabular}
					\caption[]{Squeezing of hypergraphs with single hyperedge containing all vertices}
					\label{squeezing_of_connected_hypergraphs_with_single_hyperedge_table}
				\end{table}
				
				\begin{figure}
					\centering
					\begin{subfigure}{0.15\textwidth}
						\begin{tikzpicture}[scale = .5]
							\draw[fill] (0, 0) circle [radius = 4pt];
							\node[above right] at (0, 0) {$0$};
							\draw[fill] (0, 3) circle [radius = 4pt];
							\node[below right] at (0, 3) {$1$};
							\draw[fill] (3, 0) circle [radius = 4pt];
							\node[above left] at (3, 0) {$3$};
							\draw[fill] (3, 3) circle [radius = 4pt];
							\node[below left] at (3, 3) {$2$};
							\draw [rounded corners] (-0.5, 0) -- (-0.5, 3.5) -- (3.5, 3.5) -- (3.5, -.5) -- (-0.5,-0.5) -- (-0.5, 0);
						\end{tikzpicture}
						\caption{$d = 4$, and $S_{\hat{P}} = -0.2238$}
					\end{subfigure} 
					~
					\begin{subfigure}{0.15\textwidth}
						\begin{tikzpicture}[scale = .5]
						\draw[fill] (0, 0) circle [radius = 4pt];
						\node[above right] at (0, 0) {$0$};
						\draw[fill] (0, 3) circle [radius = 4pt];
						\node[below right] at (0, 3) {$1$};
						\draw[fill] (3, 0) circle [radius = 4pt];
						\node[above left] at (3, 0) {$3$};
						\draw[fill] (3, 3) circle [radius = 4pt];
						\node[below left] at (3, 3) {$2$};
						\draw[fill] (1.5, 1.5) circle [radius = 4pt];
						\node[below] at (1.5, 1.5) {$4$};
						\draw [rounded corners] (-0.5, 0) -- (-0.5, 3.5) -- (3.5, 3.5) -- (3.5, -.5) -- (-0.5,-0.5) -- (-0.5, 0);
						\end{tikzpicture}
						\caption{$d = 5$, and $S_{\hat{P}} = -0.6817$}
					\end{subfigure} 
					~
					\begin{subfigure}{0.15\textwidth}
						\begin{tikzpicture}[scale = .5]
						\draw[fill] (0, 0) circle [radius = 4pt];
						\node[above right] at (0, 0) {$0$};
						\draw[fill] (0, 3) circle [radius = 4pt];
						\node[below right] at (0, 3) {$2$};
						\draw[fill] (3, 3) circle [radius = 4pt];
						\node[below left] at (3, 3) {$3$};
						\draw[fill] (3, 0) circle [radius = 4pt];
						\node[left] at (3, 1.5) {$4$};
						\draw[fill] (3, 1.5) circle [radius = 4pt];
						\node[above left] at (3, 0) {$5$};
						\draw[fill] (0, 1.5) circle [radius = 4pt];
						\node[right] at (0, 1.5) {$1$};
						\draw [rounded corners] (-0.5, 0) -- (-0.5, 3.5) -- (3.5, 3.5) -- (3.5, -.5) -- (-0.5,-0.5) -- (-0.5, 0);
						\end{tikzpicture}
						\caption{$d = 6$, and $S_{\hat{P}} = -0.8686$}
					\end{subfigure} 
					~
					\begin{subfigure}{0.15\textwidth}
						\begin{tikzpicture}[scale = .5]
						\draw[fill] (0, 0) circle [radius = 4pt];
						\node[above right] at (0, 0) {$0$};
						\draw[fill] (0, 3) circle [radius = 4pt];
						\node[below right] at (0, 3) {$2$};
						\draw[fill] (3, 3) circle [radius = 4pt];
						\node[below left] at (3, 3) {$3$};
						\draw[fill] (3, 0) circle [radius = 4pt];
						\node[left] at (3, 1.5) {$4$};
						\draw[fill] (3, 1.5) circle [radius = 4pt];
						\node[above left] at (3, 0) {$5$};
						\draw[fill] (0, 1.5) circle [radius = 4pt];
						\node[right] at (0, 1.5) {$1$};
						\draw[fill] (1.5, 1.5) circle [radius = 4pt];
						\node[above] at (1.5, 1.5) {$6$};
						\draw [rounded corners] (-0.5, 0) -- (-0.5, 3.5) -- (3.5, 3.5) -- (3.5, -.5) -- (-0.5,-0.5) -- (-0.5, 0);
						\end{tikzpicture}
						\caption{$d = 7$, and $S_{\hat{P}} = -0.9449$}
					\end{subfigure}
					~
					\begin{subfigure}{0.15\textwidth}
						\begin{tikzpicture}[scale = .5]
							\draw[fill] (0, 0) circle [radius = 4pt];
							\node[above right] at (0, 0) {$0$};
							\draw[fill] (0, 1.5) circle [radius = 4pt];
							\node[right] at (0, 1.5) {$1$};
							\draw[fill] (0, 3) circle [radius = 4pt];
							\node[below right] at (0, 3) {$2$};
							\draw[fill] (1.5, 3) circle [radius = 4pt];
							\node[below] at (1.5, 3) {$3$};
							\draw[fill] (3, 3) circle [radius = 4pt];
							\node[below left] at (3, 3) {$4$};
							\draw[fill] (3, 1.5) circle [radius = 4pt];
							\node[left] at (3, 1.5) {$5$};
							\node[above left] at (3, 0) {$6$};
							\draw[fill] (3, 0) circle [radius = 4pt];
							\node[above] at (1.5, 0) {$7$};
							\draw[fill] (1.5, 0) circle [radius = 4pt];
							\draw [rounded corners] (-0.5, 0) -- (-0.5, 3.5) -- (3.5, 3.5) -- (3.5, -.5) -- (-0.5,-0.5) -- (-0.5, 0);
						\end{tikzpicture}
						\caption{$d = 8$, and $S_{\hat{P}} = -0.9764$}
					\end{subfigure}
					~
					\begin{subfigure}{0.15\textwidth}
						\begin{tikzpicture}[scale = .5]
							\draw[fill] (0, 0) circle [radius = 4pt];
							\node[above right] at (0, 0) {$0$};
							\draw[fill] (0, 1.5) circle [radius = 4pt];
							\node[right] at (0, 1.5) {$1$};
							\draw[fill] (0, 3) circle [radius = 4pt];
							\node[below right] at (0, 3) {$2$};
							\draw[fill] (1.5, 3) circle [radius = 4pt];
							\node[below] at (1.5, 3) {$3$};
							\draw[fill] (3, 3) circle [radius = 4pt];
							\node[below left] at (3, 3) {$4$};
							\draw[fill] (3, 1.5) circle [radius = 4pt];
							\node[left] at (3, 1.5) {$5$};
							\node[above left] at (3, 0) {$6$};
							\draw[fill] (3, 0) circle [radius = 4pt];
							\node[above] at (1.5, 0) {$7$};
							\draw[fill] (1.5, 0) circle [radius = 4pt];
							\node[above right] at (1.5, 1.5) {$8$};
							\draw[fill] (1.5, 1.5) circle [radius = 4pt];
							\draw [rounded corners] (-0.5, 0) -- (-0.5, 3.5) -- (3.5, 3.5) -- (3.5, -.5) -- (-0.5,-0.5) -- (-0.5, 0);
						\end{tikzpicture}
						\caption{$d = 9$, and $S_{\hat{P}} = -0.9898$}
					\end{subfigure}
					\caption{Hypergraphs with exactly one hyperedge containing all vertices. The border indicates the hyperedge and the numbered dots are the vertices.}
					\label{connected_hypergraphs_with_single_edges}
				\end{figure}
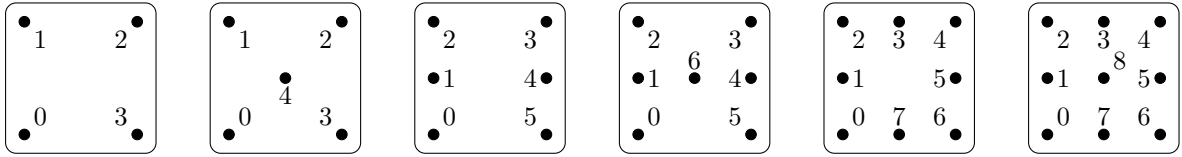
			\item 
				A $k$-graph is a hypergraph containing the hyperedges with $k$ vertices only. There is no state showing phase squeezing corresponding to connected $3$-graphs with $4$ vertices. For $d \geq 5$ we have connected $(d - 1)$-graphs with phase squeezing. The maximum and minimum values of squeezing in $(d - 1)$-graphs are mentioned in the table \ref{maximum_and_minimum_squeezing_for_(d-1)_graphs_table} and plotted in the figure \ref{maximum_and_minimum_squeezing_for_(d-1)_graphs_figure}. The hypergraphs with the maximum and the minimum degree of phase squeezing are depicted in figure \ref{connected_d_1_hypergraph_with_maximum_squeezing} and \ref{connected_d_1_hypergraph_with_miniimum_squeezing}, respectively.
				
				\begin{table}
\begin{tabular}{| r | l | p{5cm} | l | p{5cm} |}
						\hline
						$d$ & $\max_{\ket{G}}\{S_{\hat{P}}\}$ & Combination of hyperedges obtaining $\max_{\ket{G}}\{S_{\hat{P}}\}$ & $\min_{\ket{G}}\{S_{\hat{P}}\}$ & Combination of hyperedges obtaining $\min_{\ket{G}}\{S_{\hat{P}}\}$ \\
						\hline 
						$5$ & $- 0.0265$ & [(0, 1, 2, 3), (0, 1, 2, 4), (0, 1, 3, 4), (0, 2, 3, 4)] & $- 0.4968$ & $[(0, 1, 2, 3), (0, 1, 2, 4), (0, 1, 3, 4)]$ \\
						\hline
						$6$ & $-0.0066$ & [(0, 1, 2, 3, 4), (0, 1, 3, 4, 5), (0, 2, 3, 4, 5), (1, 2, 3, 4, 5)] & $- 0.7862$  & [(0, 1, 2, 3, 4), (0, 1, 2, 3, 5), (0, 1, 2, 4, 5)] \\
						\hline
						$7$ & $-0.1013$ & [(0, 1, 2, 3, 5, 6), (0, 1, 3, 4, 5, 6), (0, 2, 3, 4, 5, 6), (1, 2, 3, 4, 5, 6)] & $- 0.9113$ & [(0, 1, 2, 3, 4, 5), (0, 1, 2, 3, 4, 6), (0, 1, 2, 3, 5, 6)] \\
						\hline
						$8$ & $-0.1273$ & [(0, 1, 2, 4, 5, 6, 7), (1, 2, 3, 4, 5, 6, 7)] & $-0.9633$ & [(0, 1, 2, 3, 4, 5, 6), (0, 1, 2, 3, 4, 6, 7)] \\
						\hline
						$9$ & $-0.5749$ & [(0, 1, 2, 4, 5, 6, 7, 8), (1, 2, 3, 4, 5, 6, 7, 8)] & $- 0.9851$ & [(0, 1, 2, 3, 4, 5, 6, 7), (0, 1, 2, 3, 4, 5, 7, 8)] \\
						\hline
						$10$ & $-0.3705$ & [(0, 1, 3, 4, 5, 6, 7, 8, 9), (0, 2, 3, 4, 5, 6, 7, 8, 9)] & $-0.9937$ & [(0, 1, 2, 3, 4, 5, 6, 7, 8), (0, 1, 2, 3, 4, 5, 6, 8, 9)] \\
						\hline 
						$11$ & $-0.6754$ & [(0, 1, 3, 4, 5, 6, 7, 8, 9, 10), (0, 2, 3, 4, 5, 6, 7, 8, 9, 10)] & $-0.9973$ & [(0, 1, 2, 3, 4, 5, 6, 7, 8, 9), (0, 1, 2, 3, 4, 5, 6, 7, 9, 10)] \\
						\hline
						$12$ & $-0.8281$ & [(0, 1, 3, 4, 5, 6, 7, 8, 9, 10, 11), (0, 2, 3, 4, 5, 6, 7, 8, 9, 10, 11)] & $-0.9988$ & [(0, 1, 2, 3, 4, 5, 6, 7, 8, 9, 11), (0, 1, 2, 3, 4, 5, 6, 7, 8, 10, 11), (0, 1, 2, 3, 4, 5, 6, 7, 9, 10, 11)]\\
						\hline 
					\end{tabular}
					\caption[]{Maximum and minimum squeezing for $(d-1)$-graphs}
					\label{maximum_and_minimum_squeezing_for_(d-1)_graphs_table}
				\end{table}
			
				\begin{figure}
					\centering
					\begin{subfigure}{0.15\textwidth}
						%
						\caption{$d = 5$, and $\max_{\ket{G}}\{S_{\hat{P}}\} = - 0.0265$}
					\end{subfigure} 
					\hspace{1cm}
					\begin{subfigure}{0.15\textwidth}
						%
						\caption{$d = 6$, and $\max_{\ket{G}}\{S_{\hat{P}}\} = -0.0066$}
					\end{subfigure} 
					\hspace{1cm}
					\begin{subfigure}{0.15\textwidth}
						%
						\caption{$d = 7$, and $\max_{\ket{G}}\{S_{\hat{P}}\} = -0.1013$}
					\end{subfigure} 
					\hspace{1cm}
					\begin{subfigure}{0.15\textwidth}
						%
						\caption{$d = 8$, and $\max_{\ket{G}}\{S_{\hat{P}}\} =-0.1273 $}
					\end{subfigure}
					\\
					\begin{subfigure}{0.15\textwidth}
						%
						\caption{$d = 9$, and $\max_{\ket{G}}\{S_{\hat{P}}\} =-0.5749 $}
					\end{subfigure}
					\hspace{1cm}
					\begin{subfigure}{0.15\textwidth}
						%
						\caption{$d = 10$, and $\max_{\ket{G}}\{S_{\hat{P}}\} =-0.3705 $}
					\end{subfigure}
					\hspace{1cm}
					\begin{subfigure}{0.15\textwidth}
						%
						\caption{$d = 11$, and $\max_{\ket{G}}\{S_{\hat{P}}\} = $}
					\end{subfigure}
					\hspace{1cm}
					\begin{subfigure}{0.15\textwidth}
						%
						\caption{$d = 12$, and $\max_{\ket{G}}\{S_{\hat{P}}\} = $}
					\end{subfigure}
					\caption{Connected $(d-1)$-graph with maximum squeezing. Here the vertices are represented by the horizontal lines and the hyperedges are represented by the vertical lines. The inclusion of a vertex in a hyperedge is indicated by a bullet ($\bullet$) at the intersection.}
					\label{connected_d_1_hypergraph_with_maximum_squeezing} 
				\end{figure}

				\begin{figure}
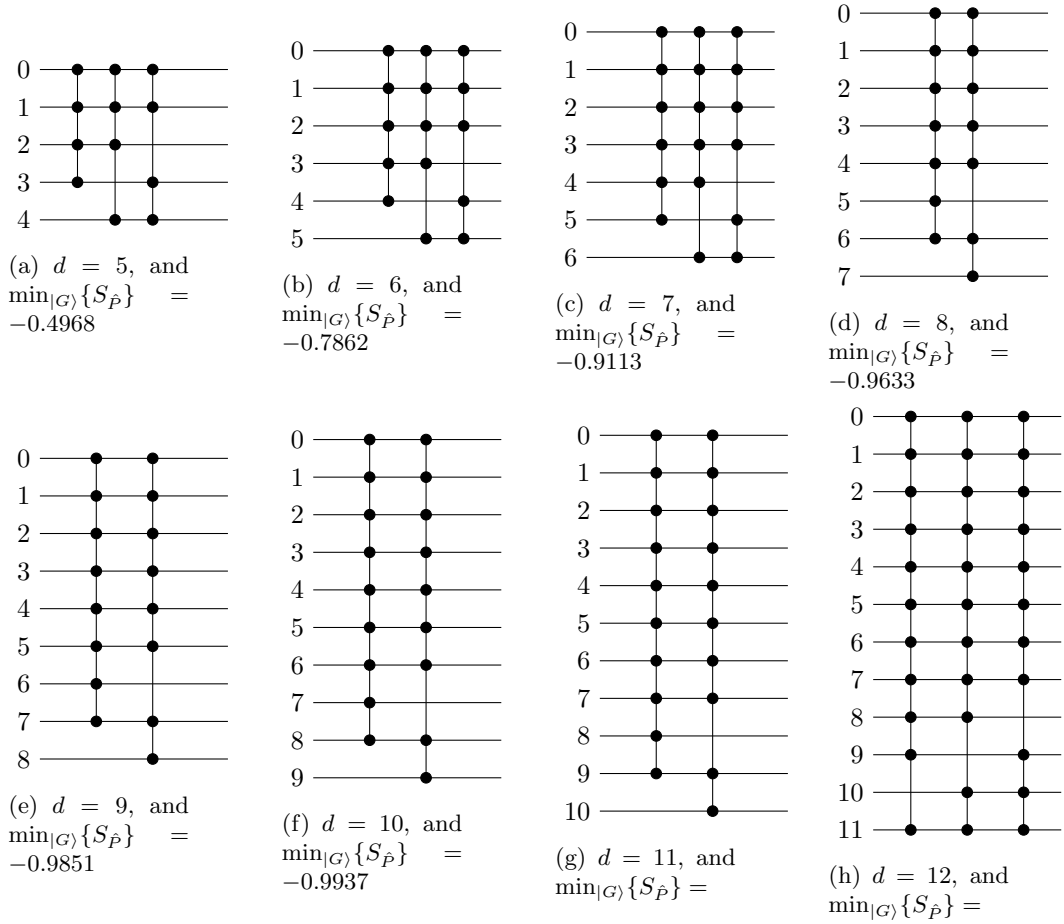

					\centering
					\begin{subfigure}{0.15\textwidth}
						%
						\caption{$d = 5$, and $\min_{\ket{G}}\{S_{\hat{P}}\} =- 0.4968 $}
					\end{subfigure} 
					\hspace{1cm}
					\begin{subfigure}{0.15\textwidth}
						%
						\caption{$d = 6$, and $\min_{\ket{G}}\{S_{\hat{P}}\} =- 0.7862 $}
					\end{subfigure} 
					\hspace{1cm}
					\begin{subfigure}{0.15\textwidth}
						%
						\caption{$d = 7$, and $\min_{\ket{G}}\{S_{\hat{P}}\} =- 0.9113 $}
					\end{subfigure} 
					\hspace{1cm}
					\begin{subfigure}{0.15\textwidth}
						%
						\caption{$d = 8$, and $\min_{\ket{G}}\{S_{\hat{P}}\} =-0.9633 $}
					\end{subfigure}
					\\
					\begin{subfigure}{0.15\textwidth}
						%
						\caption{$d = 10$, and $\min_{\ket{G}}\{S_{\hat{P}}\} =-0.9937 $}
					\end{subfigure}
					\hspace{1cm}
					\begin{subfigure}{0.15\textwidth}
						%
						\caption{$d = 11$, and $\min_{\ket{G}}\{S_{\hat{P}}\} = $}
					\end{subfigure}
					\hspace{1cm}
					\begin{subfigure}{0.15\textwidth}
						%
						\caption{$d = 12$, and $\min_{\ket{G}}\{S_{\hat{P}}\} = $}
					\end{subfigure}
					\caption{Connected $(d-1)$-graph with minimum squeezing. Here the vertices are represented by the horizontal lines and the hyperedges are represented by the vertical lines. The inclusion of a vertex in a hyperedge is indicated by a bullet ($\bullet$) at the intersection.}
					\label{connected_d_1_hypergraph_with_miniimum_squeezing}
				\end{figure}

			\item 
				The complete $k$-graphs with $d$ vertices exhibit phase squeezing when $d \geq 5$ and $k \geq 4$. The hypergraph state corresponding to the complete graph has no phase squeezing when $d < 11$. For different values of $d$ and $k$ the values of phase-squeezing are shown in table \ref{squeezing_in_complete_k_graph_table}.
				\begin{table}
					\begin{tabular}{| r | l | l | l | l | l | l | l | l | l | l |}
						\hline
						$d$ & $k = 2$ & $k = 3$ & $k = 4$ & $k = 5$ & $k = 6$ & $k = 7$ & $k = 8$ & $k = 9$ & $k = 10$ & $k = 11$ \\
						\hline 
						$5$ & & & $-0.401$ & $-0.6817$ & & & &  & & \\
						\hline
						$6$ & & $-0.5061$ & $-0.664$ & $-0.5307$ & $-0.8686$ & & &  & & \\
						\hline
						$7$ & & $-0.2925$ & $-0.8166$ & $- 0.6357$ & $-0.8636$ & $-0.9449$ & &  & & \\
						\hline
						$8$ & & & $-0.873$ & $-0.6821$ & $-0.8753$ & $-0.9253$ & $-0.9764$ &  & & \\
						\hline
						$9$ & & $-0.6502$ & $-0.9085$ & $-0.4676$ & $-0.8921$ & $-0.9178$ & $-0.9736$ & $-0.9898$ & & \\
						\hline
						$10$ & & $-0.8093$ & $-0.9366$ & $-0.8181$ & $-0.8715$ & $-0.8964$ & $-0.9769$ & $-0.9868$ & $-0.9955$ & \\
						\hline 
						$11$ & $-0.1274$ & $-0.8312$ & $-0.959$ & $-0.9602$ & $-0.712$ & & $-0.9832$ & $-0.988$ & $-0.9949$ & $-0.998$ \\
						\hline 
					\end{tabular}
					\caption[]{Squeezing in complete $k$-graph}
					\label{squeezing_in_complete_k_graph_table}
				\end{table}
			\end{enumerate}

	\section{Agarwal-Tara criterion}\label{discussions_on_Agarwal_Tara_criterion} 
		
		The Agarwal-Tara criterion is a well-known criterion of non-classicality of quantum states \cite{agarwal1992nonclassical}. It is stronger than many other criteria, such as, the determination of squeezing and sub-Poissonian photon statistics, because it may reveal nonclassicality even when the other criteria fail. Using the conventional notations used in literature we mention that for any classical probability distribution the matrix 
		\begin{equation}
			m^{(n)} = \begin{bmatrix}
			1 & m_1 & \dots  & m_{n-1} \\
			m_1 & m_2 & \dots &\\
			\vdots & \vdots & \ddots & \vdots \\
			m_{n-1} &  & \dots & m_{2n-2} \\
			\end{bmatrix}
		\end{equation}
		is positive definite given any value of $n = 1, 2, 3, \dots$, where $m_k = \braket{(a^\dagger)^k a^k}$. The existence of negative eigenvalues of $m^{(n)}$ is a witness of non-classicality. Following this observation, a measure of non-classicality is represented by
		\begin{equation}\label{Agarwal_Tara_criterion}
			A_n = \frac{\det m^{(n)}}{\det \mu^{(n)}- \det  m^{(n)}} <  0, ~\text{where}~ 
			\mu^{(n)} = \begin{bmatrix}
			1 & \mu_1 & \dots  &\mu_{n-1} \\
			\mu_1 & \mu_2 & \dots & \\
			\vdots & \vdots & \ddots & \vdots\\
			\mu_{n-1} &  & \dots & \mu_{2n-2} \\
			\end{bmatrix}
		\end{equation}
		contains the moments of number operator $\mu_k = \braket{\hat{N}^k} = \braket{(a^\dagger a)^k}$.
		
		In appendix \ref{appendix_B}, we explicitly construct the expressions of $m_k$ and $\mu_k$ for different values of $k$. We observe that these values depend only on the number of the vertices $d$ of the hypergraph. Below we summarize our numerical findings:
		\begin{enumerate}
			\item 
				The non-classicality measure $A_2 < 0$ for $d = 2$. For higher values of $d$ we have $A_2 > 0$. In table \ref{Agarwal_Tara_measure_of nonclassicality_for_A_2} the values of $A_2$ for $d = 2$ and $3$ are tabulated.
				\begin{table} 
					\begin{center}
						\begin{tabular}{| p{1 cm} | p{1 cm} | p{1.25 cm} | p{1.25 cm} | p{1 cm} | p{1 cm} | p{1 cm} | p{1 cm} |}
							\hline
							& $m_1 = \braket{a^\dagger a}$ & $m_2 = \braket{(a^\dagger)^2 a^2}$ & $\mu_1 =  \braket{\hat{N}}$ & $\mu_2 = \braket{\hat{N}^2}$ & $\det m^{(2)}$ & $\det \mu^{(2)}$ & $A_2$ \\
							\hline
							$d = 2$ &1.5 &2 &1.5  &3.5  & -0.25 & 1.25 & -0.166 \\
							\hline
							$d = 3$ &3.5 & 14&3.5  & 17.5 & 1.75 & 5.25 & 0.5 \\
							\hline 
						\end{tabular}
						\caption{Agarwal-Tara measure of non-classicality for $A_2$}
						\label{Agarwal_Tara_measure_of nonclassicality_for_A_2} 
					\end{center} 
				\end{table} 
			\item 
				For $d = 2$ the number states $\ket{0}, \ket{1}, \ket{2}$ and $\ket{3}$ forms the basis of the hypergraph states. Hence, we can not determine the moments of the number operators $\braket{\hat{N}^k}$ for $k > 3$. Therefore, we calculate $A_3$ when $d \geq 3$. We find that $A_3 < 0$ for $d = 3$ and $d = 4$. For the larger values of $d$ non-classicality can not be determined by $A_3$ which assumes positive value. Table \ref{Agarwal_Tara_measure_of nonclassicality_for_A_3} presents the values of $A_3$ for $d = 3, 4$ and $5$.
				\begin{table} 
					\begin{center} 
						\begin{tabular}{| p{1 cm} | p{1.2 cm} | p{1.2 cm} | p{1.2 cm} | p{1.5 cm} | p{1.5 cm} | p{1.5 cm} | p{1.2 cm} |}
							\hline
							& $m_3 = \braket{(a^\dagger)^3 a^3}$ & $m_4 = \braket{(a^\dagger)^4 a^4}$ & $\mu_3 = \braket{\hat{N}^3}$ & $\mu_4 = \braket{\hat{N}^4}$ & $\det m^{(3)}$ & $\det \mu^{(3)}$ & $A_3$ \\
							\hline
							$d = 3$ &52.5 &168 & 98 & 584.5 &-61.2499  &110.25  & -0.3571 \\
							\hline 
							$d = 4$ &682.5 &6552 & 760 &11144.5  & -2091.25 & 7586.25 & -0.2160  \\
							\hline  
							$d = 5$ &6742.5 &151032 & 7068 &526984.5  & 77600.749 & 494194.25 & 0.1862 \\
							\hline  
						\end{tabular}
						\caption{Agarwal-Tara measure of non-classicality for $A_3$}
						\label{Agarwal_Tara_measure_of nonclassicality_for_A_3} 
					\end{center}
				\end{table} 
			\item 
				In a similar fashion, we calculate the values of $A_4$ for $d = 3, 4$ and $5$ and collect them in the table \ref{Agarwal_Tara_measure_of nonclassicality_for_A_4}. 
				\begin{table} 
					\begin{center} 
						\begin{tabular}{| p{1 cm} | p{1.3 cm} | p{1.3 cm} | p{1.3 cm} | p{1.6 cm} | p{1.6 cm} | p{2 cm} | p{1.3 cm} |}
							\hline
							& $m_5 = \braket{(a^\dagger)^5 a^5}$ & $m_6 = \braket{(a^\dagger)^6 a^6}$ & $\mu_5 = \braket{\hat{N}^5}$ & $\mu_6 = \braket{\hat{N}^6}$ & $\det m^{(4)}$ & $\det \mu^{(4)}$ & $A_4$ \\
							\hline
							$d = 3$ &420 &720 & 3526 & 23102.5 & 12405393 &187211.06  & -1.0153 \\
							\hline 
							$d = 4$ &60060 &514800 &190792.5  &2028032.5  &$3.01293\times 10^{11}$  &$1.5322\times 10^{11}$  & -2.0348 \\
							\hline  
							$d = 5$ &3398220 & 75731760&5081768  & $1.371\times 10^{18}$  &$-2.6151\times 10^{18}$
							& $-6.6556\times10^{16}$ & -1.0261 \\
							\hline  
						\end{tabular}
						\caption{Agarwal-Tara measure of non-classicality for $A_4$}
						\label{Agarwal_Tara_measure_of nonclassicality_for_A_4} 
					\end{center}
				\end{table} 
		\end{enumerate}
		The above numerical calculations suggest that the Agarwal-Tara criterion for non-classicality is satisfied by the quantum hypergraph states. Depending on the number of vertices $d$ in the hypergraphs $A_n < 0$ for $n = 2, 3, 4$ and $d = 2, 3, 4, 5$.

	\section{Coherence}\label{coherence}
		
		In the literature, quantum coherence is studied via the measures; the relative entropy of coherence, and the $l_1$ norm of coherence. Given a density matrix $\rho$, we define the matrix $\rho_{\diag}$ by deleting all off-diagonal elements. The relative entropy of coherence is defined by \cite{baumgratz2014quantifying}
		\begin{equation}
			\mathcal{C}_{rel.ent}(\rho) = S(\rho_{\diag}) - S(\rho),
		\end{equation}
		where $S(\rho)$ is the von-Neumann entropy. The $l_1$ norm of coherence is given by $\mathcal{C}_{l_1}(\rho) = \sum_{i, j, i \neq j}|\rho_{ij}|$.
		
\begin{figure}
			\centering
			\begin{subfigure}{0.45\textwidth}
				\includegraphics[scale = .45]{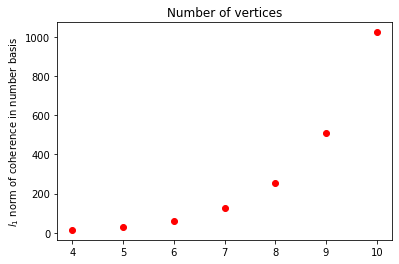}
				\caption{$l_1$ norm of coherence for any hypergraph in number basis} 
				\label{$l_1$ norm of coherence for hypergraphs in number basis}
			\end{subfigure} 
			\hspace{1cm}
			\begin{subfigure}{0.45\textwidth}
				\includegraphics[scale = .45]{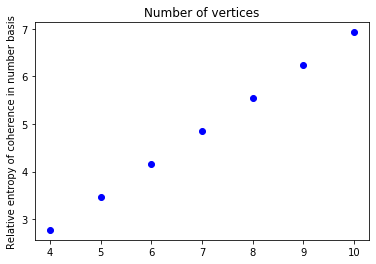}
				\caption{Relative entropy of coherence for any hypergraph in number basis} 
				\label{Relative entropy coherence for hypergraphs in number basis}
			\end{subfigure} 
			\caption{Coherence in number basis. The number of vertices are plotted in the $x$ axis.}
		\end{figure}

		The quantum hypergraph state $\ket{G}$ mentioned in equation (\ref{optical_hypergraph_state}) is also represented by the density matrix $\rho_G = \ket{G} \bra{G} = \frac{1}{2^d}(\rho_{i,j})_{2^d \times 2^d}$, where $\rho_{i,j} \in \{1, -1\}$. In other words, the absolute value of the entries of $\rho_G$ is $\frac{1}{2^d}$, when it is expressed in number basis. As the hypergraph state is a pure state the von-Neumann entropy $S(\rho_G) = 0$. Therefore, the relative entropy of coherence of quantum hypergraph states in number basis \cite{dutta2019permutation} is
		\begin{equation}
			\mathcal{C}_{rel.ent}(\rho_G) = S(\rho_{\diag}) = - 2^d \times \frac{1}{2^d} \times \log \left(\frac{1}{2^d}\right) = d\log(2).
		\end{equation} 
		In addition, the $l_1$ norm of coherence for any quantum hypergraph state $C_{l1}(\rho_G) = 2^d-1$. Coherence in $l_1$ norm and relative entropy  are plotted in figure \ref{$l_1$ norm of coherence for hypergraphs in number basis} and \ref{Relative entropy coherence for hypergraphs in number basis}, respectively.
		
		We also calculate coherence of the hypergraph states in phase basis. We find that different classes of hypergraph states have different values of coherence. Here, we study coherence of states for $(d - 1)$-graphs.
		\begin{table}
			\centering
			\begin{tabular}{| c | c | c | c | c | c | c | c |}
				\hline
				$d$ & 4 & 5 & 6 & 7 & 8 & 9 & 10 \\
				\hline
				Coherence in $l_1$ norm & 15 & 31 & 63 & 127 & 255 & 511 & 1023\\
				\hline  
			\end{tabular}
			\caption[]{Coherence in $l_1$ norm in number basis for $(d-1)$-graphs}
			\label{Coherence in $l_1$ norm in number basis for $(d-1)$-graphs}
		\end{table}
		\begin{table}
			\centering 
			\begin{tabular}{| c | c | c | c | c | c | c | c |}
				\hline 
					$d$ & 4 & 5 & 6 & 7 & 8 & 9 & 10 \\
				\hline
					Coherence in entropy & $2.772$ & $3.465$ & $4.158$ & $4.852$ & $5.545$ & $6.238$ & $6.931$\\
				\hline 
			\end{tabular}
			\caption[]{Coherence in entropy in number basis for $(d-1)$-graphs}
			\label{Coherence in entropy in number basis for $(d-1)$-graphs}
		\end{table}

		\begin{table}
\begin{tabular}{| r | l | p{6.5cm} | l | p{3.5cm} |}
				\hline
				$d$ & $\max_{\ket{G}}$ $C_e$ & Combination of hyperedges obtaining $\max_{\ket{G}}$  $C_e$ & $\min_{\ket{G}}$ $C_e$& Combination of hyperedges obtaining $\min_{\ket{G}}$ $C_e$ \\
				\hline 
				$4$ & $2.4889$ & (0, 1, 2), (0, 1, 3), (0, 2, 3), (1, 2, 3) & $1.709$ & $(0, 2, 3), (1, 2, 3)$ \\
				\hline 
				$5$ & $2.5315$ & (0, 1, 2, 3), (0, 1, 2, 4), (0, 1, 3, 4), (0, 2, 3, 4),(1, 2, 3, 4) & $1.2645$ & $(0, 2, 3, 4), (1, 2, 3, 4)$ \\
				\hline
				$6$ & $1.9539$ & (0, 1, 2, 3, 4), (0, 1, 2, 3, 5), (0, 1, 2, 4, 5), (0, 1, 3, 4, 5), (0, 2, 3, 4, 5), (1, 2, 3, 4, 5) & $0.8317$  & (0, 1, 2, 3, 4), (1, 2, 3, 4, 5) \\
				\hline
				$7$ & $1.579$ &(0, 1, 2, 3, 4, 5), (0, 1, 2, 3, 4, 6), (0, 1, 2, 3, 5, 6), (0, 1, 2, 4, 5, 6), (0, 1, 3, 4, 5, 6), (0, 2, 3, 4, 5, 6), (1, 2, 3, 4, 5, 6) & $0.496$ & (0, 1, 2, 4, 5, 6), (1, 2, 3, 4, 5, 6) \\
				\hline
				$8$ & $0.9944$ & (0, 1, 2, 3, 4, 5, 6), (0, 1, 2, 3, 4, 5, 7), (0, 1, 2, 3, 4, 6, 7), (0, 1, 2, 3, 5, 6, 7), (0, 1, 2, 4, 5, 6, 7), (0, 1, 3, 4, 5, 6, 7), (0, 2, 3, 4, 5, 6, 7), (1, 2, 3, 4, 5, 6, 7) & $ 0.2513$ & (0, 1, 3, 4, 5, 6, 7), (1, 2, 3, 4, 5, 6, 7) \\
				\hline
				$9$ &0.6774&(0, 1, 2, 3, 4, 5, 6, 7), (0, 1, 2, 3, 4, 5, 6, 8), (0, 1, 2, 3, 4, 5, 7, 8), (0, 1, 2, 3, 4, 6, 7, 8), (0, 1, 2, 3, 5, 6, 7, 8), (0, 1, 2, 4, 5, 6, 7, 8), (0, 1, 3, 4, 5, 6, 7, 8), (0, 2, 3, 4, 5, 6, 7, 8), (1, 2, 3, 4, 5, 6, 7, 8) &0.1324  & (0, 2, 3, 4, 5, 6, 7, 8), (1, 2, 3, 4, 5, 6, 7, 8) \\
				\hline
				$10$ & 0.3135 &(0, 1, 2, 3, 4, 5, 6, 7,8), (0, 1, 2, 3, 4, 5, 6, 8,9), (0, 1, 2, 3, 4, 5, 7, 8,9), (0, 1, 2, 3, 4, 6, 7, 8,9), (0, 1, 2, 3, 5, 6, 7, 8,9), (0, 1, 2, 4, 5, 6, 7, 8,9), (0, 1, 3, 4, 5, 6, 7, 8,9), (0, 2, 3, 4, 5, 6, 7, 8,9), (1, 2, 3, 4, 5, 6, 7, 8,9),(0, 1, 2, 3, 4, 5, 6, 7,9)  & 0.0078&(0, 2, 3, 4, 5, 6, 7, 8, 9), (1, 2, 3, 4, 5, 6, 7, 8, 9) \\
				\hline
			\end{tabular}
			\caption[]{Maximum and minimum coherence in entropy in phase basis for $(d-1)$-graphs}
			\label{Maximum and minimum coherence in entropy in phase basis for $(d-1)$-graphs}
		\end{table}
		\begin{table}
			\begin{tabular}{| r | l | p{6.5cm} | l | p{3.5cm} |}
				\hline
				$d$ & $\max_{\ket{G}}$ $C_{l1}$ & Combination of hyperedges obtaining $\max_{\ket{G}}$  $C_{l1}$ & $\min_{\ket{G}}$ $C_{l1}$& Combination of hyperedges obtaining $\min_{\ket{G}}$ $C_{l1}$ \\
				\hline 
				$4$ & $12.8646$ & (0, 1, 2), (0, 1, 3), (0, 2, 3), (1, 2, 3) & 7.4926 & $(0, 2, 3), (1, 2, 3)$ \\
				\hline 
				$5$ & 19.4148 & (0, 1, 2, 3), (0, 1, 2, 4), (0, 1, 3, 4), (0, 2, 3, 4),(1, 2, 3, 4) &9.0122 & $(0, 2, 3, 4), (1, 2, 3, 4)$ \\
				\hline
				$6$ & $25.6414$ & (0, 1, 2, 3, 4), (0, 1, 2, 3, 5), (0, 1, 2, 4, 5), (0, 1, 3, 4, 5), (0, 2, 3, 4, 5), (1, 2, 3, 4, 5) & $10.8189$  & (0, 2, 3, 4,5), (1, 2, 3, 4, 5) \\
				\hline
				$7$ & 31.4497 &(0, 1, 2, 3, 4, 5), (0, 1, 2, 3, 4, 6), (0, 1, 2, 3, 5, 6), (0, 1, 2, 4, 5, 6), (0, 1, 3, 4, 5, 6), (0, 2, 3, 4, 5, 6), (1, 2, 3, 4, 5, 6) & 10.234 & (0, 2, 3, 4, 5, 6), (1, 2, 3, 4, 5, 6) \\
				\hline
				$8$ & $35.805$ & (0, 1, 2, 3, 4, 5, 6), (0, 1, 2, 3, 4, 5, 7), (0, 1, 2, 3, 4, 6, 7), (0, 1, 2, 3, 5, 6, 7), (0, 1, 2, 4, 5, 6, 7), (0, 1, 3, 4, 5, 6, 7), (0, 2, 3, 4, 5, 6, 7), (1, 2, 3, 4, 5, 6, 7) & $11.4444$ & (0, 2, 3, 4, 5, 6, 7), (1, 2, 3, 4, 5, 6, 7) \\
				\hline
			\end{tabular}
			\caption[]{Maximum and minimum coherence in $l_1$ norm in phase basis for $(d-1)$-graphs}
			\label{Maximum and minimum coherence in $l_1$ norm in phase basis for $(d-1)$-graphs}
		\end{table}
		\begin{figure}
			\centering
			\begin{subfigure}{0.15\textwidth}
				%
				\caption{$d = 4$, $\max_{\ket{G}}\{C_{rel.ent}\} =2.4889 $}
			\end{subfigure} 
			\hspace{1.5cm}
			\begin{subfigure}{0.15\textwidth}
				%
				\caption{$d = 5$, $\max_{\ket{G}}\{C_{rel.ent}\} =2.5315 $}
			\end{subfigure} 
			\hspace{1.5cm}
			\begin{subfigure}{0.15\textwidth}
				%
				\caption{$d = 6$, $\max_{\ket{G}}\{C_{rel.ent}\} =1.9539 $}
			\end{subfigure}
            \\
			\begin{subfigure}{0.15\textwidth}
				%
				\caption{$d = 7$, $\max_{\ket{G}}\{C_{rel.ent}\} = 1.579$}
			\end{subfigure} 
			\hspace{2.5cm}
			\begin{subfigure}{0.15\textwidth}
				%
				\caption{$d = 8$, and $\max_{\ket{G}}\{C_{rel.ent}\} = 0.9944$}
			\end{subfigure}
			\caption{Connected $(d-1)$-graph with maximum relative entropy of coherence calculated in phase basis. Here the vertices are represented by the horizontal lines and the hyperedges are represented by the vertical lines. The inclusion of a vertex in a hyperedge is indicated by a bullet ($\bullet$) at the intersection.}
			\label{connected_d_1_hypergraph_with_maximum_entropy}
		\end{figure}
		\begin{figure}
			\centering
			\begin{subfigure}{0.15\textwidth}
				%
				\caption{$d = 4$, and $\min_{\ket{G}}\{C_{rel.ent.}\} = 1.709$}
			\end{subfigure} 
			\hspace{1cm}
			\begin{subfigure}{0.15\textwidth}
				%
				\caption{$d = 5$, and $\min_{\ket{G}}\{C_{rel.ent}\} =1.2645 $}
			\end{subfigure} 
			\hspace{1cm}
			\begin{subfigure}{0.15\textwidth}
				%
				\caption{$d = 6$, and $\min_{\ket{G}}\{C_{rel.ent}\} =0.8317 $}
			\end{subfigure} 
			\hspace{1cm}
			\begin{subfigure}{0.15\textwidth}
				%
				\caption{$d = 7$, and $\min_{\ket{G}}\{C_{rel.ent}\} = 0.496$}
			\end{subfigure} 
			\hspace{1cm}
			\begin{subfigure}{0.15\textwidth}
				%
				\caption{$d = 8$, $\min_{\ket{G}}\{C_{rel.ent}\} = 0.2513$}
			\end{subfigure}
			\hspace{1.5cm}
			\begin{subfigure}{0.15\textwidth}
				%
				\caption{$d = 9$, and $\min_{\ket{G}}\{C_{rel.ent}\} =0.1324 $}
			\end{subfigure}
			\hspace{1.5cm}
			\begin{subfigure}{0.15\textwidth}
				%
				\caption{$d = 10$, and $\min_{\ket{G}}\{C_{rel.ent}\} =0.0078 $}
			\end{subfigure}
			\caption{Connected $(d-1)$-graph with minimum relative entropy coherence calculated in phase basis. Here the vertices are represented by the horizontal lines and the hyperedges are represented by the vertical lines. The inclusion of a vertex in a hyperedge is indicated by a bullet ($\bullet$) at the intersection.}
			\label{connected_d_1_hypergraph_with_miniimum_relative_entropy}
		\end{figure}
		\begin{figure}
			\centering
			\begin{subfigure}{0.15\textwidth}
				%
				\caption{$d = 4$, and $\min_{\ket{G}}\{C_{l1}\} = 7.4926$}
			\end{subfigure} 
			\hspace{.75cm}
			\begin{subfigure}{0.15\textwidth}
				%
				\caption{$d = 5$, and $\min_{\ket{G}}\{C_{l1}\} =9.0122 $}
			\end{subfigure} 
			\hspace{.75cm}
			\begin{subfigure}{0.15\textwidth}
				%
				\caption{$d = 6$, and $\min_{\ket{G}}\{C_{l1}\} =10.8189 $}
			\end{subfigure} 
			\hspace{.75cm}
			\begin{subfigure}{0.15\textwidth}
				%
				\caption{$d = 7$, and $\min_{\ket{G}}\{C_{l1}\} = 10.234$}
			\end{subfigure} 
			\hspace{.75cm}
			\begin{subfigure}{0.15\textwidth}
				%
				\caption{$d = 8$, $\min_{\ket{G}}\{C_{l1}\} = 11.4444$}
			\end{subfigure}
			\caption{Connected $(d-1)$-graph with minimum $l_1$ norm coherence, calculated in phase basis. Here the vertices are represented by the horizontal lines and the hyperedges are represented by the vertical lines. The inclusion of a vertex in a hyperedge is indicated by a bullet ($\bullet$) at the intersection.}
			\label{connected_d_1_hypergraph_with_miniimum_l1}
		\end{figure}
		\begin{figure} 
			\centering 
			\begin{subfigure}{.4 \textwidth}
				\includegraphics[scale = .45]{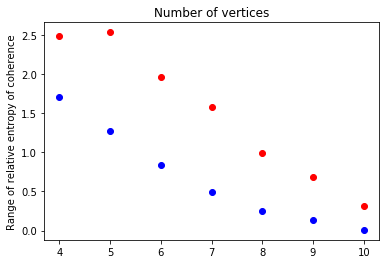}
				\caption{Maximum and minimum values of relative entropy coherence for $(d - 1)$-graphs. (Red and blue points represent the minimum and maximum.)} 
				\label{Maximum and minimum values of relative entropy of coherence for $(d - 1)$-graphs}
			\end{subfigure}
			\hspace{1cm}
			\begin{subfigure}{.4 \textwidth}
				\includegraphics[scale = .45]{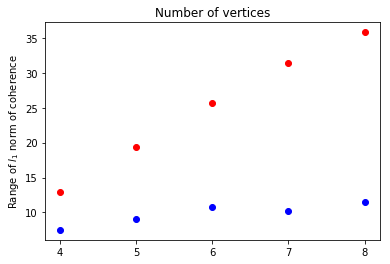}
				\caption{Maximum and minimum values of $l_1$ norm of coherence for $(d - 1)$-graphs. (Red and blue points represent the minimum and maximum.)} 
				\label{maximum_and_minimum_l1_for_(d-1)_graphs_figure}
			\end{subfigure}
			\caption{Range of coherence for $(d-1)$ graphs. The number of vertices $d$ is plotted in the $x$-axis.}
		\end{figure} 
		\begin{enumerate}
			\item 
				An interesting observation is that if the hypergraph contains all possible combination of $(d - 1)$ hyperedges then the corresponding state has the maximum value of relative entropy as well as the maximum value of $l_1$ norm. These hypergraphs are plotted in the figure \ref{connected_d_1_hypergraph_with_maximum_entropy}.
			\item  
				The maximum and minimum values of relative entropy of coherence in phase basis of $(d - 1)$-graphs are listed in the table \ref{Maximum and minimum coherence in entropy in phase basis for $(d-1)$-graphs} and plotted in the figure \ref{Maximum and minimum values of relative entropy of coherence for $(d - 1)$-graphs}. The hypergraphs with minimum coherence of entropy are depicted in figure \ref{connected_d_1_hypergraph_with_miniimum_relative_entropy}.
			\item 
				The maximum and minimum values of $l_1$ norm of  $(d - 1)$ graphs in phase basis are plotted in figure \ref{maximum_and_minimum_l1_for_(d-1)_graphs_figure}. The hypergraphs with minimum coherence of $l_1$ norms are depicted in figure  \ref{connected_d_1_hypergraph_with_miniimum_l1}.  
		\end{enumerate}
		Since coherence is observed in phase basis, this observation is consistent with the earlier observation of squeezing, a quantum feature, being observed in the phase quadrature.

	\section{Concluding remarks}
	
		Given any hypergraph with $d$ vertices, there is a quantum hypergraph state in $2^d$ dimensional Hilbert space $\mathcal{H}^{2^d}$. We studied the non-classical properties of these states. Their number-phase uncertainty relations were examined. It was observed that though there is no squeezing in the number quadrature for any hypergraph, there are states squeezed in the phase quadrature. We chose a number of hypergraphs and numerically compute their phase squeezing, which include the connected hypergraph with single hyperedge, complete $(d - 1)$-graphs. In case of the connected hypergraphs with single hyperedge the degree of squeezing is close to $-1$ when $d$ increases. We also establish that the Agarwal-Tara criterion for non-classicality holds for quantum hypergraph states when number of vertices $d \leq 5$. Our numerical observations may help an interested reader, in future, to produce a general statement in this regard applicable to all hypergraph states. Coherence, an important facet of quantumness, was also studied in different basis using various measures for the relevant states. It was seen that coherence exhibits non-trivial behavior in phase basis.
		
		An interested reader may attempt further works in this direction. The concepts  of squeezing and general uncertainty relations in  continuous variable quantum states have been investigated in the literature \cite{agarwal2005inseparability}. It can be extended to the discrete qubit based entangled states, considering the examples of highly entangled hypergraph states. For this purpose, the appropriate oscillator algebra in the finite dimensional Hilbert space should be constructed with a hypergraph based network of multi-party oscillators. This naturally leads to analogs  of squeezing and uncertainty relations, which assist in characterizing the underlying entanglement. This is in parallel to the use of Peres-Horodecki PPT criterion of the finite dimensional Hilbert space to the Gaussian entangled states. Now, an entanglement measure through the uncertainty relations can be proposed. One may extend it for the non-Gaussian states, based on higher order uncertainties, originating from $SU(2)$ and $SU(1,1)$ group theoretical descriptions of quantum optical entangled states.

	\section*{Acknowledgment}
	
		SD and RS have equal contribution to this work. RS is thankful to Ministry of Human Resource and Development, Government of India for a doctoral fellowship via IISER Kolkata. SB and PKP acknowledges support from Interdisciplinary Cyber Physical Systems (ICPS) programme of the Department of Science and Technology (DST), India through Grant No.: DST/ICPS/QuEST/Theme-1/2019/6. SB also acknowledges support from Interdisciplinary Research Platform- Quantum Information and Computation (IDRP-QIC) at IIT Jodhpur.

	\appendix
	
	\section{Essential concepts of linear algebra}\label{appendix_A}

		\noindent \textbf{Hermitian and skew-Hermitian matrix:} A matrx $\hat{A}$ is called Hermitan if $\hat{A}^\dagger = \hat{A}$, where $\hat{A}^\dagger$ is the conjugate transpose of $\hat{A}$. Also, a matrix $\hat{A}$ is skew-Hermitan if $\hat{A}^\dagger = -\hat{A}$. Note that, if $\hat{A}$ is skew-Hermitian, then $\iota \hat{A}$ is Hermitan, where $\iota$ is the complex identity.
			
		\begin{lemma}
			\textbf{Gershgorin circle theorem}: Let $\hat{A} = (a_{i,j})_{n \times n}$ be a complex square matrix and $R_i = \sum_{j \neq i, j = 1}^n |a_{i, j}|$. Also, $D(a_{i,i}, R_i) \subset \mathbb{C}$ is a closed dice centered at $a_{i,i}$ of radius $R_i$. Then, every eigenvalue of $A$ lies within at least one of the discs $D(a_{i,j}, R_i)$. In other words for any eigenvalue $\lambda$ there is an $i$, such that \cite{horn2012matrix},
			$$|\lambda - a_{i, i}| \leq \sum_{j \neq i, j = 1}^n |a_{i, j}|.$$
		\end{lemma}
	
		\noindent \textbf{Rayleigh quotient:} Given any Hermitian matrix $\hat{A}$ and a non-zero vector $x$, the Rayleigh quotient is defined by \cite{horn2012matrix}
		\begin{equation}
		R(\hat{A}, x) = \frac{x^\dagger \hat{A} x}{x^\dagger x}. 
		\end{equation}
		If $x$ represents a quantum state $\ket{x}$, then $x^\dagger x = \braket{x|x} = 1$, and the Rayleigh quotient is $R(\hat{A}, \ket{x}) = \braket{x| \hat{A} | x}$. If $\lambda_{\min}$ and $\lambda_{\max}$ be the minimum and maximum eigenvalues of $\hat{A}$, then $\lambda_{\min} \leq R(\hat{A}, \ket{x}) \leq \lambda_{\max}$, for any quantum state $\ket{x}$.
		
		\noindent \textbf{Circulant matrix:} A circulant matrix \cite{gray2006toeplitz} $C$ or order $n$ has the following form
		\begin{equation}
			C = \begin{bmatrix}
				c_0 & c_{n - 1} & \dots & c_2 & c_1 \\
				c_1 & c_0 & c_{n - 1} &  \dots & c_2 \\
				\vdots & c_1 & c_0 & \ddots & \vdots \\
				c_{n - 2} & \ddots & \ddots & \ddots & c_{n - 1}\\
				c_{n - 1} & c_{n - 2} & \dots & c_1 & c_0
			\end{bmatrix}.
		\end{equation}
		Recall the phase operator $\hat{P}$ in equation (\ref{phase_operator_in_number_state}), which is a matrix of order $2^d$. If $s = |k - l|$ for some $k$ and $l$ then 
		\begin{equation}
			c_s = \frac{2 \pi}{4^d} P_{k, l} = \frac{2 \pi}{4^d} \sum_{r = 0}^{2^d - 1} r \exp \left(\frac{2 \pi \iota r |k - l|}{2^d} \right) = \frac{2 \pi}{4^d} \sum_{r = 0}^{2^d - 1} r \exp \left(\frac{2 \pi \iota r s}{2^d} \right).
		\end{equation}
		Also, it is easy to check that $c_{2^d - s} = \overline{c_s}$. Therefore, $\hat{P}$ is a Hermitian circulant matrix.
		
		\noindent \textbf{Toeplitz matrix:} An $n \times n$ Toeplitz matrix \cite{gray2006toeplitz} $A$ has the form
		\begin{equation}
			A = \begin{bmatrix}a_{0}&a_{-1}&a_{-2}&\cdots &\cdots &a_{-(n-1)}\\a_{1}&a_{0}&a_{-1}&\ddots &&\vdots \\a_{2}&a_{1}&\ddots &\ddots &\ddots &\vdots \\\vdots &\ddots &\ddots &\ddots &a_{-1}&a_{-2}\\\vdots &&\ddots &a_{1}&a_{0}&a_{-1}\\a_{n-1}&\cdots &\cdots &a_{2}&a_{1}&a_{0}\end{bmatrix}
		\end{equation}
		The equation (\ref{commutation_operator_in_number_basis}) suggests that the elements of $[\hat{N}, \hat{P}]$ can be expressed as
		\begin{equation}
			a_s = \frac{2 \pi s}{2^d} \sum_{r = 0}^{2^d - 1} r \exp \left(\frac{2 \pi \iota rs}{2^d}\right),
		\end{equation}
		where $s = (k - l)$. It is easy to verify that $a_{-s} = - \overline{a_s}$. Therefore, matrix $[\hat{N}, \hat{P}]$ is a skew-Hermitian Toeplitz matrix.

	\section{Calculations for Agarwal-Tara criterion}\label{appendix_B}
	
		Here, we calculate the higher order moments of the number operator $\hat{N}$ and verify the Agarwal-Tara criterion of non-classicality mentioned in equation (\ref{Agarwal_Tara_criterion}) for different values of $n$. We observe that these calculations depend on the number of vertices in the hypergraph but independent of the distribution of the hyperedges. The calculations are as follows.
		
		Applying an annihilation operator on $\ket{\psi_0} = \ket{G}$, mentioned in equation (\ref{optical_hypergraph_state}) we obtain an unnormalized state vector,
		\begin{equation}\label{first_anihilation_1}
			a\ket{\psi_0} = \frac{1}{\sqrt{2^d}} \sum_{i = 0}^{2^d - 1} (-1)^{f(i)} a\ket{i} = \frac{1}{\sqrt{2^d}} \sum_{i = 1}^{2^d - 1} (-1)^{f(i)} \sqrt{i}\ket{i - 1}.
		\end{equation}
		The normalization factor is given by $\frac{1}{\sqrt{W_1}}$, where
		\begin{equation}
			W_1 = \sum_{i = 1}^{2^d - 1}\left[\frac{(-1)^{f(i)} \sqrt{i}}{\sqrt{2^d}}\right]^2 = \frac{1}{2^d} \sum_{i = 1}^{2^d - 1} i = \frac{2^d (2^d - 1)}{2. 2^d} = \frac{(2^d - 1)}{2}.
		\end{equation}
		Normalizing the state in equation (\ref{first_anihilation_1}) we find 
		\begin{equation}\label{1_st_normalize_state}
			\ket{\psi_1} = \sqrt{\frac{2}{2^d(2^d - 1)}} \sum_{i = 1}^{2^d - 1} (-1)^{f(i)} \sqrt{i}\ket{i - 1}.
		\end{equation}
		Symbolically, we can write $a\ket{\psi_0} = \sqrt{W_1} \ket{\psi_1}$. Applying the annihilation operator on $\ket{\psi_1}$ we have
		\begin{equation}
		a \ket{\psi_1} = \sqrt{\frac{2}{2^d(2^d - 1)}} \sum_{i = 1}^{2^d - 2} (-1)^{f(i + 1)} \sqrt{i + 1} \sqrt{i} \ket{i - 1}.
		\end{equation}
		Here, the coefficient of $\ket{i - 1}$ is $\sqrt{\frac{2}{2^d(2^d - 1)}} (-1)^{f(i + 1)} \sqrt{i + 1} \sqrt{i}$. Hence, the normalization factor will be given by $\frac{1}{\sqrt{W_2}}$, where  
		\begin{equation}
		\begin{split}
		W_2 = & \sum_{i = 1}^{2^d - 2} \left[\sqrt{\frac{2}{2^d(2^d - 1)}} (-1)^{f(i + 1)} \sqrt{i + 1} \sqrt{i}\right]^2 = \frac{2}{2^d(2^d - 1)} \sum_{i = 1}^{2^d - 2} i(i + 1) \\
		= & \frac{2}{2^d(2^d - 1)} \frac{1}{3}(2^d - 2) (2^d - 1) (2^d) = \frac{2(2^d - 2)}{3}.
		\end{split}
		\end{equation}
		After normalization the state is
		\begin{equation}\label{2_nd_normalized_state}
			\begin{split}
				\ket{\psi_2} = & \sqrt{\frac{3}{2 (2^d - 2) }} \sqrt{\frac{2}{2^d(2^d - 1)}} \sum_{i = 1}^{2^d - 2} (-1)^{f(i + 1)} \sqrt{i + 1} \sqrt{i} \ket{i - 1} \\
				= & \sqrt{\frac{3}{2^d (2^d - 1) (2^d - 2)}} \sum_{i = 0}^{2^d - 3} (-1)^{f(i + 2)} \sqrt{i + 2} \sqrt{i + 1} \ket{i}.
			\end{split}
		\end{equation}
		Now, equation (\ref{2_nd_normalized_state}) indicates $a^2\ket{\psi_0} = a \sqrt{W_1} \ket{\psi_1} = \sqrt{W_1W_2}\ket{\psi_2}$. In general, for $0 < k < (2^d - 1)$, we can prove that
		\begin{equation}\label{annihilation_operator_in_general}
			\begin{split} 
				& a^k \ket{\psi_0} = \sqrt{W_1 W_2 \dots W_k} \ket{\psi_k}, \\
				\text{where}~ & \ket{\psi_k} = \sqrt{\frac{k+1}{^{2^d}P_{(k + 1)}}} \sum_{i = 0}^{2^d - (k+1)} (-1)^{f(i +k)} \sqrt{^{i + k}P_k} \ket{i}, ~\text{and}~ W_k = \frac{k(2^d - k)}{k + 1}.
			\end{split} 
		\end{equation}
		Here, we denote $^{2^d}P_k = \frac{2^d!}{(2^d - k)!} = 2^d(2^d - 1) (2^d - 2) \dots (2^d - k + 1)$. Several numerical values of $W_k$ depending on $d$ are mentioned in the table below.
		
		\begin{center} 
			\begin{tabular}{| p{1 cm} | p{1 cm} | p{1 cm} | p{1 cm} | p{1 cm} | p{1 cm} | p{1 cm} |}
				\hline
				& $W_1$ & $W_2$ & $W_3$ & $W_4$ & $W_5$ & $W_6$ \\
				\hline
				$d = 2$ &$\frac{3}{2}$  &$\frac{4}{3}$  &$\frac{3}{4}$  &  &  &\\
				\hline
				$d = 3$ &$\frac{7}{2}$  &4  &$\frac{15}{4}$  & $\frac{16}{5}$ &$\frac{5}{2}$  &$\frac{12}{7}$\\
				\hline 
				$d = 4$ & $\frac{15}{2}$ & $\frac{28}{3}$ & $\frac{39}{4}$ &  $\frac{48}{5}$& $\frac{55}{6}$ &$\frac{60}{7}$\\
				\hline  
				$d = 5$ & $\frac{31}{2}$ &20 & $\frac{87}{4}$ & $\frac{112}{5}$ & $\frac{45}{2}$ & $\frac{156}{7}$\\
				\hline  
			\end{tabular}\\ 
		\end{center} 
	
		Note that, the calculation of $W_k$ for $k \geq 2$ needs the notion of sum of products of consecutive integers \cite{summation_method}. Given two dummy indices $r$ and $i$ we define $f_r(i) = i(i + 1)(i + 2) \dots (i + r - 1)$ and $F_r(i) = \frac{1}{r + 1}i(i + 1)(i + 2) \dots (i + r)$. Clearly, $F_r(0) = 0$, and
		\begin{equation}
			\begin{split}
				F_r(i) - F_r(i - 1) = & \frac{1}{r + 1}i(i + 1)(i + 2) \dots (i + r) - \frac{1}{r + 1}(i - 1)i(i + 1) \dots (i + r - 1) \\
				= & \frac{1}{r + 1} i (i + 1) \dots (i + r - 1) [(i + r) - (i - 1)] = f_r(i).
			\end{split}
		\end{equation}
		Therefore, for any integer $n$ with $1 \leq n \leq (2^d - 1)$ we have
		\begin{equation}
			\sum_{i = 1}^n f_r(i) = F_r(n) = \frac{1}{r + 1}n(n + 1)(n + 2) \dots (n + r)
		\end{equation}
		For $r = 1$ and $n = (2^d - 1)$ we have
		\begin{equation}
			\sum_{i = 1}^{2^d - 1} f_1(i) = \sum_{i = 1}^{2^d - 1} i = \frac{(2^d - 1)2^d}{2},
		\end{equation}
		and for $r = 2$ and $n = (2^d - 2)$ we have
		\begin{equation}
			\sum_{i = 1}^{2^d - 2} f_2(i) = \sum_{i = 1}^{2^d - 2} i(i + 1) = \frac{(2^d - 2)2^d(2^d + 1)}{3}.
		\end{equation}
		
		Note that, the dimension of the state $\ket{\psi_k}$ is $2^d$ for all $k$. Also, the coefficients of the vectors $\ket{i}$ for $i > 2^d - (k+1)$ in $\ket{\psi_k}$ are $0$. Hence, they are excluded from the above equation. Considering conjugate transpose on both sides of equation (\ref{annihilation_operator_in_general}), we have 
		\begin{equation}\label{creation_operator_in_general}
			\begin{split} 
				& \bra{\psi_0} {a^\dagger}^k = \sqrt{W_1 W_2 \dots W_k} \bra{\psi_k}\\
				\text{where}~ & \bra{\psi_k} = \sqrt{\frac{k + 1}{^{2^d}P_{(k + 1)}}} \sum_{i = 0}^{2^d - (k + 1)} (-1)^{f(i + k)} \sqrt{^{i + k}P_k} \bra{i}.
			\end{split} 
		\end{equation}
		Equations (\ref{annihilation_operator_in_general}) and (\ref{creation_operator_in_general}) together indicate
		\begin{equation}\label{average_a_dagger_k_a_k}
			m_k = \braket{{a^\dagger}^k  {a^k}} = \braket{\psi_0 | {a^\dagger}^k  {a^k} | \psi_0} = W_1W_2 \dots W_k \braket{\psi_k | \psi_k} = W_1W_2 \dots W_k,
		\end{equation}
		since, $\braket{\psi_k | \psi_k} = 1$. For $k = 1$ , we have $\braket{a^\dagger a} = W_1$ and for $k = 2$ we have $\braket{(a^2)^\dagger a^2} = W_1W_2$.
		
		Next, we calculate the values of $\mu_k = \braket{\hat{N}^k} = \braket{(a^\dagger a)^k}$. Recall that, in finite dimensions the relation between the creation and the annihilation operators are given by equation (\ref{a_a_dagger_commutation}). Thus
		\begin{equation}
			\begin{split}
				& \hat{N}^2 = (a^\dagger a)^2 = a^\dagger a a^\dagger a = a^\dagger (I + a^\dagger a-2^d\ket{2^d-1} \bra{2^d-1}) a = a^\dagger a + (a^\dagger)^2 a^2 - 0 \\
				\text{or}~ & \braket{\hat{N}^2} = \braket{a^\dagger a} + \braket{(a^\dagger)^2 a^2} = W_1 + W_1W_2.
			\end{split}
		\end{equation}
		
		Now we take up the Agarwal-Tara criterion for checking non-classicality. Putting $n = 2$ in equation (\ref{Agarwal_Tara_criterion}) and applying equation (\ref{average_a_dagger_k_a_k}), we have 
		\begin{equation}
			\begin{split}
				& \det m^{(2)} = \begin{vmatrix} 1 & \langle (a)^\dagger a \rangle \\ \langle (a)^\dagger a \rangle & \langle (a^2)^\dagger a^2 \rangle \end{vmatrix} = \begin{vmatrix} 1 & W_1 \\ W_1 & W_1W_2 \end{vmatrix} = W_1W_2 -W_1^2, \\
				\text{and}~ & \det \mu^{(2)} = \begin{vmatrix} 1 & \langle a^\dagger a \rangle \\ \langle a^\dagger a \rangle & \langle (a^\dagger a)^2 \rangle \end{vmatrix} = = \begin{vmatrix} 1 & W_1 \\ W_1 & W_1+W_1W_2 \end{vmatrix}  = W_1+W_1W_2 -W_1^2.
			\end{split}
		\end{equation}
		Hence,
		\begin{equation}
			\begin{split}
				A_2 = \frac{\det m^{(2)}}{\det \mu^{(2)}- \det  m^{(2)}} = \frac{W_1W_2 -W_1^2}{W_1+W_1W_2 -W_1^2-(W_1W_2 -W_1^2)} = W_2 - W_1 = \frac{2(2^d - 2)}{3} - \frac{(2^d - 1)}{2}.
			\end{split}
		\end{equation}
		For $d = 2$ and $3$ we calculate the values of $\det m^{(2)}, \det \mu^{(2)}$ and $A_2$ and collect them in table \ref{Agarwal_Tara_measure_of nonclassicality_for_A_2}.
	
		For calculating $A_3$ we need higher values of $\braket{\hat{N}^3}$ and $\braket{\hat{N}^4}$. Making use of the commutation relation between $a$ and $a^\dagger$, equation (\ref{a_a_dagger_commutation}), we have
		\begin{equation}
			\begin{split}
				(a^\dagger)^2 a^2 . a^\dagger a & = (a^\dagger)^2 a (aa^\dagger) a = (a^\dagger)^2 a (I + a^\dagger a- 2^d\ket{2^d-1} \bra{2^d-1}) a\\ & = (a^\dagger)^2 a^2 + (a^\dagger)^2 a a^\dagger a^2-0 = (a^\dagger)^2 a^2 + (a^\dagger)^2 (I + a^\dagger a-2^d\ket{2^d-1} \bra{2^d-1}) a^2 \\& =(a^\dagger)^2 a^2 + (a^\dagger)^2 a^2 + (a^\dagger)^3 a^3 -0 = 2 (a^\dagger)^2 a^2 + (a^\dagger)^3 a^3,\\
				\text{and}~ (a^\dagger)^3 a^3 . a^\dagger a & = (a^\dagger)^3 a^2 (a a^\dagger) a = (a^\dagger)^3 a^2 ( I + a^\dagger a-2^d\ket{2^d-1} \bra{2^d-1}) a \\ &= (a^\dagger)^3 a^3 + (a^\dagger)^3 a^2 a^\dagger a^2 -0
				= (a^\dagger)^3 a^3 + (a^\dagger)^3 a(a a^\dagger) a^2 \\ &=(a^\dagger)^3 a^3 + (a^\dagger)^3 a(I + a^\dagger a- 2^d\ket{2^d-1} \bra{2^d-1}) a^2\\
				&= (a^\dagger)^3 a^3 + (a^\dagger)^3 a^3 + (a^\dagger)^3 a a^\dagger a^3-0\\& = 2 (a^\dagger)^3 a^3 + (a^\dagger)^3 (I + a^\dagger a-2^d\ket{2^d-1} \bra{2^d-1}) a^3 \\
				& = 2 (a^\dagger)^3 a^3 + (a^\dagger)^3 a^3 + (a^\dagger)^4 a^4 -0= 3 (a^\dagger)^3 a^3 + (a^\dagger)^4 a^4.
			\end{split} 
		\end{equation}
		Continuing in this fashion we find
		\begin{equation}\label{general_higher_equation}
		(a^\dagger)^k a^k . a^\dagger a = k (a^\dagger)^k a^k + (a^\dagger)^{k + 1} a^{k + 1},
		\end{equation}
		for $k = 1, 2, 3, \dots $.	
		
		Applying this relation it can be seen that
		\begin{equation}\label{average_N_cube}
			\begin{split}
				\hat{N^3} & =(a^\dagger a)^3=\hat{N^2} a^\dagger a =
				(a^\dagger a+ a^{2\dagger} a^2 )(a^\dagger a)=a^\dagger a a^\dagger a+a^{2\dagger} a^2 a^\dagger a\\
				& = a^\dagger a+ a^{2\dagger} a^2+ 2a^{2\dagger} a^2+ a^{3\dagger} a^3=
				a^\dagger a+ 3a^{2\dagger} a^2+a^{3\dagger} a^3\\
				\text{or}~ \braket{\hat{N^3}} & = W_1+ 3W_1W_2+W_1W_2W_3,\\
				\text{and}~ \hat{N^4} & =(a^\dagger a)^4=\hat{N^3} a^\dagger a = (a^\dagger a+ 3a^{2\dagger} a^2+a^{3\dagger} a^3)(a^\dagger a)\\
				& = a^\dagger a a^\dagger a+ 3a^{2\dagger} a^2 a^\dagger a+ a^{3\dagger} a^3 a^\dagger a =a^\dagger a+ a^{2\dagger} a^2+ 6a^{2\dagger} a^2+3a^{3\dagger} a^3+ 3a^{3\dagger} a^3+a^{4\dagger} a^4\\
				& = a^\dagger a+ 7a^{2\dagger} a^2+6a^{3\dagger} a^3 + a^{4\dagger} a^4\\
				\text{or}~ \braket{\hat{N^4}} & = W_1+ 7W_1W_2+6W_1W_2W_3+W_1W_2W_3W_4. 
		\end{split}
		\end{equation}
		Putting these values in the Agarwal-Tara criterion for $n = 3$ we find
		\begin{equation}
			\begin{split}
				\det m^{(3)} & = \begin{vmatrix} 1 & m_1 & m_2 \\ m_1 & m_2 & m_3 \\ m_2 & m_3 & m_4
				\end{vmatrix} = \begin{vmatrix} 1 & \langle a^\dagger a \rangle & \langle (a^2)^\dagger a^2 \rangle \\ \langle a^\dagger a \rangle & \langle (a^2)^\dagger a^2 \rangle & \langle (a^3)^\dagger a^3 \rangle \\ \langle (a^2)^\dagger a^2 \rangle & \langle (a^3)^\dagger a^3 \rangle & \langle (a^4)^\dagger a^4 \rangle \end{vmatrix} = \begin{vmatrix}
				1 & W_1 & W_1W_2\\
				W_1 & W_1W_2 & W_1W_2W_3\\
				W_1W_2 & W_1W_2W_3 &W_1W_2W_3W_4
				\end{vmatrix}\\
				& = W_1^2W_2W_3(W_2-W_1)(W_4-W_2)+W_1^3W_2^2(W_3-W_2).\\
				\text{and}~ \det \mu^{(3)} = & \begin{vmatrix} 1 & \mu_1 & \mu_2 \\ \mu_1 & \mu_2 & \mu_3 \\ \mu_2 & \mu_3 & \mu_4 \end{vmatrix} = \begin{vmatrix} 1 & \langle a^\dagger a \rangle & \langle (a^\dagger a)^2 \rangle \\ \langle a^\dagger a \rangle & \langle (a^\dagger a)^2 \rangle & \langle (a^\dagger a)^3 \rangle \\ \langle (a^\dagger a)^2 \rangle & \langle (a^\dagger a)^3 \rangle & \langle (a^\dagger a)^4 \rangle \end{vmatrix} \\ 
				= & \begin{vmatrix}
				1 & W_1 & W_1+W_1W_2\\
				W_1 & W_1+W_1W_2 & W_1+3W_1W_2+W_1W_2W_3\\
				W_1+W_1W_2 &W_1+3W_1W_2+W_1W_2W_3 &W_1+7W_1W_2+6W_1W_2W_3+W_1W_2W_3W_4
				\end{vmatrix}\\
			\end{split}
		\end{equation}
		 $\det m^{(2)}, \det \mu^{(2)}, A_3$ for different values of $d$ are tabulated in table \ref{Agarwal_Tara_measure_of nonclassicality_for_A_3}.
	
		To calculate $A_4$ we need the expressions of $\hat{N}^5$ and $\hat{N}^6$, which are as follows:
		\begin{equation}
			\begin{split}
				\hat{N}^5 & =(a^\dagger a)^5=\hat{N^4} a^\dagger a=(a^\dagger a+ 7a^{2\dagger} a^2+6a^{3\dagger} a^3+
				a^{4\dagger} a^4)(a^\dagger a)\\
				& =a^\dagger a a^\dagger a+ 7a^{2\dagger} a^2 a^\dagger a+ 6a^{3\dagger} a^3 a^\dagger a+a^{4\dagger} a^4 a^\dagger a\\
				& = a^\dagger a+ a^{2\dagger} a^2+14a^{2\dagger} a^2+ 7a^{3\dagger} a^3+18a^{3\dagger} a^3+6a^{4\dagger} a^4+4a^{4\dagger} a^4+a^{5\dagger} a^5\\
				& =a^\dagger a+ 15a^{2\dagger} a^2+25a^{3\dagger} a^3+ 10a^{4\dagger} a^4+a^{5\dagger} a^5
			\end{split}
		\end{equation}
		\begin{equation}
			\begin{split} 
				\text{or}~ \braket{\hat{N}^5} & = W_1+ 15W_1W_2+25W_1W_2W_3+10W_1W_2W_3W_4+W_1W_2W_3W_4W_5, \\
				\text{and}~\hat{N^6} & =(a^\dagger a)^6=U^5 a^\dagger a=(a^\dagger a+ 15a^{2\dagger} a^2+25a^{3\dagger} a^3+ 10a^{4\dagger} a^4+a^{5\dagger} a^5)(a^\dagger a)\\
				& = a^\dagger a a^\dagger a+ 15a^{2\dagger} a^2 a^\dagger a+25a^{3\dagger} a^3 a^\dagger a+10a^{4\dagger} a^4 a^\dagger a+ a^{5\dagger} a^5 a^\dagger a\\
				& = a^\dagger a+ a^{2\dagger} a^2+30a^{2\dagger} a^2+15a^{3\dagger} a^3+75a^{3\dagger} a^3+25a^{4\dagger} a^4\\
				& + 40a^{4\dagger} a^4+10a^{5\dagger} a^5+5a^{5\dagger} a^5+a^{6\dagger} a^6\\
				& = a^\dagger a+31a^{2\dagger} a^2+90a^{3\dagger} a^3+65a^{4\dagger} a^4+15a^{5\dagger} a^5+a^{5\dagger} a^5\\
				\text{or}~ \braket{\hat{N}^6} & = W_1+ 31W_1W_2+90W_1W_2W_3+65W_1W_2W_3W_4+15W_1W_2W_3W_4W_5+W_1W_2W_3W_4W_5 W_6 
			\end{split}
		\end{equation}
		These expressions allow us to calculate the values of $A_4$ for different values of $d$ in table \ref{Agarwal_Tara_measure_of nonclassicality_for_A_4}.

		The propagation of the coefficients in the expression of $\hat{N}^k$ for $k = 1, 2, \dots 6$ can be seen in the table below. The higher powers of $\hat{N}^k$ for larger values of $k$ will also follow a similar pattern.
		
		\begin{center}
			\begin{tabular}{| p{.5cm} | p{.5cm} | p{1.5cm} | p{1.5cm} | p{1.5cm} | p{1.5cm} | p{1.5cm} |}
				\hline
				& $a^\dagger a$ & $(a^\dagger)^2 a^2$ & $(a^\dagger)^3 a^3$ & $(a^\dagger)^4 a^4$ & $(a^\dagger)^5 a^5$ & $(a^\dagger)^6 a^6$ \\
				\hline
					$\hat{N}$ & $1$ &  & &&&\\
				\hline
					$\hat{N}^2$ & $1$ & 1 & &&&\\
				\hline 
					$\hat{N}^3$ & $1$ & $3 = 1 + 1 + 1$ & $1$ & & &\\
				\hline  
					$\hat{N}^4$ & $1$ & $7 = 1 + 3 + 1 \times 3$ & $6 = 1 + 3 + 2 \times 1$ & 1 & & \\
				\hline 
					$\hat{N}^5$ & $1$ & $ 15 = 1 + 7 + 1 \times 7$ & $25 = 7 + 6 + 2 \times 6$ & $10 = 1 + 6 + 3 \times 1$ & $1$ & \\
				\hline 
					$\hat{N}^6$ & $1$ & $31 = 1 + 15 + 1 \times 15$ & $90 = 15 + 25 + 2 \times 25$ & $65 = 25 + 10 + 3 \times 10$ & $15 = 10 + 1 + 4 \times 1$ & $1$ \\
				\hline 
			\end{tabular}
		\end{center}


\end{document}